\documentclass[sigconf]{acmart}

\AtBeginDocument{%
  }

\usepackage{balance}
\usepackage{url}
\usepackage[capitalise]{cleveref}
\usepackage{enumitem}
\usepackage{multirow}
\usepackage{booktabs}

\newcommand{\xiyuan}{\textcolor{black}}

\usepackage{color}       
\usepackage{graphicx}     

\newcommand{\Stepone}{\textcolor[RGB]{86, 128, 175}}
\newcommand{\Steptwo}{\textcolor[RGB]{96, 84, 148}}
\newcommand{\CaseOneRatio}{\textcolor{black}}

\copyrightyear{2025}
\acmYear{2025}
\acmConference[IUI '25]{30th International Conference on Intelligent User Interfaces}{March 24--27, 2025}{Cagliari, Italy}
\acmBooktitle{30th International Conference on Intelligent User Interfaces (IUI '25), March 24--27, 2025, Cagliari, Italy}\acmDOI{10.1145/3708359.3712075}
\acmISBN{979-8-4007-1306-4/25/03}

\begin{document}

\title{Prefer2SD: A Human-in-the-Loop Approach to Balancing Similarity and Diversity in In-Game Friend Recommendations}

\author{Xiyuan Wang}
\orcid{0009-0008-1839-2010}
\affiliation{%
    \institution{School of Information Science and Technology, ShanghaiTech University}
  \city{Shanghai}
  \country{China}
}
\email{wangxy7@shanghaitech.edu.cn}

\author{Ziang Li}
\orcid{0009-0005-5229-8325}
\affiliation{%
  \institution{Tongji University}
  \city{Shanghai}
  \country{China}
}
\email{ziangli@tongji.edu.cn}

\author{Sizhe Chen}
\orcid{0009-0004-5612-9935}
\affiliation{%
  \institution{Chalmers University of Technology}
  \city{Gothenburg}
  \country{Sweden}
}
\email{sizhec@chalmers.se}

\author{Xingxing Xing}
\orcid{0000-0003-3305-5158}
\affiliation{%
 \institution{UX Center, Netease Games}
 \city{Hangzhou}
 \country{China}
}
\email{xingxingxing@corp.netease.com}

\author{Wei Wan}
\orcid{0009-0002-0228-7458}
\affiliation{%
  \institution{UX Center, Netease Games}
  \city{Guangzhou}
  \country{China}
}
\email{gzwanwei@corp.netease.com}

\author{Quan Li}
\authornote{Corresponding Author.}
\orcid{0000-0003-2249-0728}
\affiliation{%
  \institution{School of Information Science and Technology, ShanghaiTech University}
  \city{Shanghai}
  \state{}
  \country{China}
  }
\email{liquan@shanghaitech.edu.cn}

\renewcommand{\shortauthors}{Wang et al.}

\begin{abstract}
In-game friend recommendations significantly impact player retention and sustained engagement in online games. Balancing similarity and diversity in recommendations is crucial for fostering stronger social bonds across diverse player groups. However, automated recommendation systems struggle to achieve this balance, especially as player preferences evolve over time. To tackle this challenge, we introduce \textit{Prefer2SD} (derived from \textbf{\underline{Prefer}}ence \underline{to} \textbf{\underline{S}}imilarity and \textbf{\underline{D}}iversity), an iterative, human-in-the-loop approach designed to optimize the similarity-diversity (SD) ratio in friend recommendations. Developed in collaboration with a local game company, \textit{Prefer2D} leverages a visual analytics system to help experts explore, analyze, and adjust friend recommendations dynamically, incorporating players' shifting preferences. The system employs interactive visualizations that enable experts to fine-tune the balance between similarity and diversity for distinct player groups. We demonstrate the efficacy of \textit{Prefer2SD} through a within-subjects study (N=$12$), a case study, and expert interviews, showcasing its ability to enhance in-game friend recommendations and offering insights for the broader field of personalized recommendation systems.
\end{abstract}





\maketitle

\section{Introduction}
\par In recommendation systems, the challenge of balancing similarity with diversity has long been a central research focus~\cite{shi2013trading,eskandanian2020using,zhang2020diversity}. In this work, we introduce the Similarity-Diversity (SD) ratio, a metric designed to quantify the trade-off between these two factors. Striking an effective balance is critical, as users tend to prefer content that aligns with their existing interests while also benefiting from exposure to novel and diverse perspectives. This balance not only enhances user experience but also helps mitigate the ``filter bubble'' effect~\cite{pariser2011filter}, \xiyuan{where repetitive exposure to similar content limits users' access to diverse new experiences}, thus encouraging exploration across a wider range of content domains.

\par The balance between similarity and diversity is equally important in friend recommendation systems within online video games, such as massively multiplayer online role-playing games (MMORPGs) and multiplayer online battle arenas (MOBAs)~\cite{yang2022large,ward2022network}. In these environments, similarity reflects players' tendencies to form connections with others who exhibit similar gaming behaviors, while diversity emphasizes the advantages of engaging with individuals possessing complementary skills or gameplay. Maintaining this balance enriches players' social experience~\cite{hinds2000choosing}, facilitating interactions with like-minded companions while also encouraging the exploration of social circles and exposure to various gameplay styles. However, despite the recognized value of this equilibrium, most current game friend recommendation systems prioritize similarity as the primary metric~\cite{Yin2022DiversityPL,Zhang2022ALF,zhang2022measuring,chang2022graphrr}, often overlooking the intricate interplay between similarity and diversity.

\par In the recommendation domain, various machine learning (ML) techniques have been proposed to dynamically adjust the SD ratio in recommendations. 
Common approaches include collaborative filtering and graph-based models, which aim to balance accuracy and diversity by optimizing user-item interactions and leveraging complex relationship structures in data~\cite{zhang2020diversity,xie2021improving,bian2011online,agarwal2013collaborative,fan2019graph,wu2022graph,gao2023survey}, ensuring that users receive both personalized and varied content. Techniques such as multi-stage optimization and graph neural networks are widely used to achieve these goals, providing flexible frameworks that adapt to different scenarios.
However, directly applying these methods to friend recommendations in online video games presents several unique challenges. \textbf{1) Rapid Environmental Changes.} The fast-paced nature of online games, driven by frequent updates and seasonal events, leads to rapid shifts in player interests, which traditional ML models struggle to adapt to. \textbf{2) Diverse Needs Across Player Phases.} Players at different stages of their gaming journey exhibit varying preferences regarding the SD ratio~\cite{vahlo2017digital}. Novices often seek connections with those of similar experience, while more experienced players tend to value diversity in recommendations~\cite{vella2020making,nardi2006strangers}. Existing recommendation methods, however, commonly adopt a ``one-size-fits-all'' approach~\cite{levit2022theory,seo2017personalized,chen2013unified,raza2024comprehensive,algarni2023systematic}, limiting their ability to cater to these diverse needs. \textbf{3) Lack of User Control.} Many friend recommendation systems, particularly in gaming communities, lack transparency and user control~\cite{Yin2022DiversityPL,riyadh2020enhancing}, which can lead to unpredictable and occasionally unfavorable outcomes, particularly in the complex and dynamic environments of online games. It raises concerns about user experience and system performance, highlighting the need for more transparent and human-in-the-loop approaches in game friend recommendation systems.

\par To bridge this gap, it is essential to involve algorithm experts \xiyuan{who design and refine friend recommendation systems for specific game environments,} integrating their domain-specific knowledge into the process. These experts have a deep understanding of the intricacies inherent in in-game social interactions, enabling them to navigate the rapidly evolving landscape and address the diverse needs of players. Their involvement brings subjective insights and detailed reasoning, enhancing recommendations to align with player expectations. Given this, developing an interactive framework with tailored visualizations for experts is essential to enhance transparency, interpretability, and recommendation effectiveness, ultimately increasing players' satisfaction. Several interactive visualization tools~\cite{bostandjiev2012tasteweights,verbert2014effect,parra2014see} have been designed to help end-users understand and participate in the recommendation process. However, these tools are primarily oriented towards end-users, offering guidance and additional input through interactions. They often lack features specifically designed for algorithm experts, which limits the experts' ability to conduct in-depth analysis, gain deeper insights, and effectively adjust the SD ratio in the recommendation process.

\par Building on the challenges and insights identified, we introduce \textit{Prefer2SD} (\textbf{\underline{Prefer}}ence \underline{to} \textbf{\underline{S}}imilarity and \textbf{\underline{D}}iversity), a human-in-the-loop solution for enhancing in-game friend recommendations. This interactive visual workflow empowers algorithm experts to optimize the SD ratio for specific player cohorts \xiyuan{they choose} through a structured two-step process. Expanding upon the traditional multi-stage recommendation pipeline (\cref{fig:mult-stage-rec}), our workflow integrates several innovations. In the candidate generation phase (\cref{fig:mult-stage-rec}(A-B)), \textit{Prefer2SD} captures players' multi-modal preferences, diversifying the friend candidate pool. Moving to the ranking stage (\cref{fig:mult-stage-rec}(B-C)), expert knowledge is employed to adjust the composition of friend candidates by carefully mediating intra- and inter-preference ratios to achieve the ideal SD ratio at the group level. \textit{Prefer2SD} was developed in collaboration with a leading local game company over a year, incorporating insights from both in-game algorithm and operation experts. It seamlessly aligns with backend friend recommendation processes while offering enhanced frontend interactivity through six coordinated views. To validate its effectiveness, we evaluate \textit{Prefer2SD} through a within-subjects study (N=$12$), a case study, and expert interviews. Results demonstrate its success in achieving desired SD ratios and improving recommendation performance for targeted player groups. The main contributions of this work are as follows:

\par 1) We present \textit{Prefer2SD}, an innovative two-step human-in-the-loop workflow that enables algorithm experts to fine-tune the SD ratio of friend recommendations for specific player groups.
\par 2) We design a highly interactive visual analytics system that aligns with the \textit{Prefer2SD} workflow, allowing algorithm experts to dynamically explore, analyze, and interpret the recommendation process. It incorporates fine-grained preference ratio adjustments to give users more precise control over recommendation outcomes. 
\par 3) We validate the efficacy of \textit{Prefer2SD} in achieving target SD ratios and enhancing recommendation performance for distinct player groups through a within-subjects study (N=$12$), a case study, and expert interviews.

\section{Background}
\subsection{About MMORPGs}
\label{sec:mmo}
\par A MMORPGis an Internet-based video game where numerous players take on their roles in a fictional world and engage in a virtual gaming environment. \xiyuan{This study focuses on MMORPGs due to their emphasis on sustained social interactions and long-term player relationships. Unlike MOBA games, where player interactions are often short-term and session-based, MMORPGs require friend recommendation systems that helps create enduring connections among players.} Although contemporary MMORPGs exhibit variations, they typically share fundamental characteristics.

\par \textbf{Social Interaction.} Social interaction is a core aspect of MMORPGs, often facilitated through \textit{\textbf{guilds}}—organized groups of players working together to achieve common goals. Guilds promote teamwork and shared goals, especially in \textit{\textbf{PvE}} settings, where cooperation is crucial to success. Working together on challenging objectives fosters a sense of camaraderie and enhances the overall social experience, as players rely on each other for strategy and support.

\par \textbf{Role-playing.} Role-playing lets players immerse themselves in the game world by adopting specific \textit{\textbf{roles or schools}} that define their abilities and responsibilities. \textit{\textbf{Avatars}}, which represent characters, are highly customizable, reflecting each player's identity and style. With numerous avatar options, players can express their individuality. Role-playing deepens engagement by combining character development with strategic gameplay and storytelling.

\par In our study, we incorporate these shared characteristics with player preferences, and categorize into three distinct parts: \textit{social}, \textit{gameplay} and \textit{avatar} aspects, which will be detailed in \cref{sec: backend}.

\subsection{Multi-stage Recommendation Systems}
\par
\begin{figure}
\centering
\includegraphics[width=\linewidth]{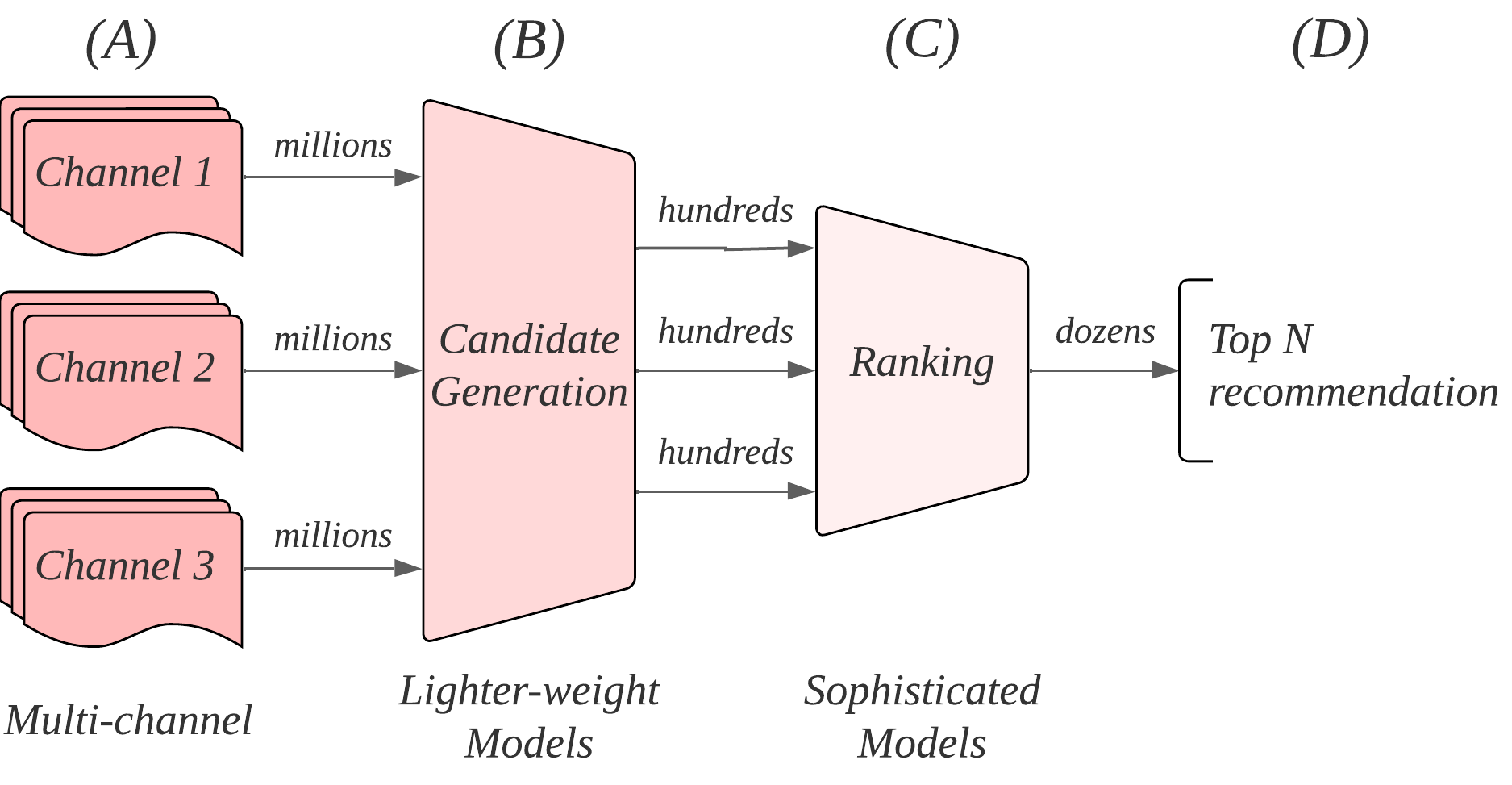}
\caption{Workflow of multi-stage recommendation systems: Illustrated is an example of multi-channel matching for candidate generation. (A) Multi-channel classifies various items into specific channels, forming an extensive recommendation pool. (B) Candidate Generation efficiently narrows down items from millions to hundreds using models designed for large data handling. (C-D) The Ranking stage then sorts items by predicted user relevance, delivering a \textit{top@N} list through sophisticated models.}
\label{fig:mult-stage-rec}
\end{figure}

 With the widespread availability of online services and content, recommendation systems play a crucial role in assisting users in discovering items of interest~\cite{zhou2019personalized}. \xiyuan{For MMORPGs, recommending friends based on player preferences is crucial to fostering long-term social interactions and enhancing player engagement within the game.} Among the various recommendation strategies, multi-stage recommendation systems have been proven to be highly effective, especially in industries~\cite{ak2022training}. These systems leverage sequential filtering and refining techniques to enhance recommendation quality, providing users with more relevant and suitable suggestions.
The architecture of multi-stage recommendation systems typically involves two main phases: \textit{candidate generation} (\cref{fig:mult-stage-rec}(B)) and \textit{ranking} (\cref{fig:mult-stage-rec}(C)). These stages enable the system to streamline the extensive item space (\cref{fig:mult-stage-rec}(A)) into a manageable set (\cref{fig:mult-stage-rec}(D)), paving the way for a more refined selection process.

\par \textbf{Candidate Generation.} In this phase (\cref{fig:mult-stage-rec}(B)), the objective is to narrow down the vast array of potential items (often in millions) to a more manageable subset of candidates (typically ranging from hundreds to thousands). Given the scale of this task, lighter-weight models such as matrix factorization, embedding-based retrieval, or shallow neural networks are commonly employed. Notably, \textit{\textbf{multi-channel matching}}~\cite{xie2020internal} proves to be an effective and common approach in the industry at this stage (\cref{fig:mult-stage-rec}(A-B)). By leveraging data from various channels (\cref{fig:mult-stage-rec}(A)) such as user browsing and purchase history, and social interactions, the system can attain a comprehensive understanding of user preferences. Each channel contributes a distinct perspective, and their integration can reflect the interests of users as a whole.

\par \textbf{Ranking.} Following the candidate generation phase, this stage (\cref{fig:mult-stage-rec}(C)) focuses on ranking the items according to their predicted relevance to the user, ultimately delivering $top@N$ recommendations (\cref{fig:mult-stage-rec}(D)) that are expected to be most appealing. Advanced models, including machine learning and deep learning architectures, are deployed in this stage, considering intricate user-item interactions, user and item metadata, context, and other nuanced factors.

\section{Related Work}

\subsection{Friend Recommendation in Games}
\par \xiyuan{Friend recommendation strategies vary significantly across game genres due to differing player interaction and social goals. This section examines the unique challenges of MMORPGs and contrasts them with the approaches typically used in MOBA and First Person Shooter (FPS) games.}

\par Friend recommendation strategies in social networks~\cite{wang2014friendbook,chen2020friend,huang2015social,cheng2019friend,bagci2016context} commonly involve constructing user graphs and utilizing various sources, such as scholars' interactivity~\cite{xu2019scholar} and users' lifestyles~\cite{wang2014friendbook}, to gauge similarity between users. While graph neural network (GNN) approaches~\cite{chang2022graphrr,wu2022graph} have attracted attention, they may not be well-suited for game friend recommendations, given the sparse nature of in-game user interactions and the challenge of designing suitable GNN architectures for various recommendation scenarios in large online game communities~\cite{li2023survey,electronics13040677,ye2021sparse}. In contrast, integrating player portraits into the friend recommendation system has shown performance improvement~\cite{watson2019study,zhang2022large}. 

\par \xiyuan{In MMORPGs, friend recommendations aim to encourage long-term relationships~\cite{ramirez2018good} by considering various and diverse player preferences, such as social behaviors and gameplay~\cite{chen2020friend}. However, in games like MOBA and FPS, the focus shifts to short-term team formation~\cite{duell2014team} for match-based gameplay. FPS games typically rely on behavior-based approaches, analyzing metrics such as win rates~\cite{peng2024application} and recent performance~\cite{claypool2007frame} to balance team compositions. MOBA games, on the other hand, emphasize real-time compatibility, often relying on communication frequency~\cite{kim2024communication} and skill-based matchmaking~\cite{pramono2018matchmaking}. While these systems are efficient for forming transient teams, they may overlook mechanisms for building lasting social connections or promoting diversity. Unlike competitive games with ELO systems~\cite{glickman1995glicko} designed for short-term compatibility, our work addresses the challenge of recommending lasting friendships, particularly in MMORPGs, where relationships often persist for extended periods.}

\subsection{Similarity and Diversity in Friend Recommendations}
\par \xiyuan{Balancing similarity and diversity has long been a central challenge in recommendation systems~\cite{eskandanian2020using,xie2021improving}. Similarity reflects users' tendencies to connect with those who share common interests or behaviors~\cite{fkih2022similarity}, while diversity encourages exposure to novel and varied experiences~\cite{kunaver2017diversity}. Achieving this balance is particularly important in gaming contexts, as it directly influences player engagement and social satisfaction.}

\par Previous studies have predominantly concentrated on suggesting similar friends, neglecting the diverse friend needs of players~\cite{mu2018survey,zhao2023fairness,noel2024improving}. Enhancing recommendation diversity can address the long tail problem~\cite{park2008long} and mitigate biased outcomes.
Current efforts to enhance recommendation diversity can be categorized into \textit{collaborative filtering-based}, \textit{content-based}, and \textit{rank-based} methods~\cite{adomavicius2011improving}. Some researchers have explored improving recommendation diversity through interface design~\cite{tsai2017leveraging}. Nevertheless, the equilibrium between similarity and diversity in game friend recommendations remains unexplored, presenting an avenue for future research in this domain. We present an interactive and dynamic framework that leverages human expertise. It is crafted to aid professionals in exploring, analyzing, and fine-tuning recommendation process to achieve an optimal equilibrium.

\par \xiyuan{In MMORPGs, diversity, as defined by the game company, refers to including players with different gameplay preferences to create balanced and engaging experiences. This study defines diversity along three main dimensions: social, gameplay, and avatar preferences. For example, in MMORPGs, where quests often require a mix of combat, support, and exploration skills, assembling teams with diverse play styles ensures a better chance of success and engagement for players~\cite{chang2014team}. This framework addresses these challenges by ensuring teams benefit from a variety of play styles rather than focusing solely on similarity, filling a gap in existing friend recommendation systems.}

\subsection{Interactive Recommendation Systems}
\par Interactive recommendation systems prioritize user-centered control and interactivity, fostering a comprehensive understanding of the recommendation process. This approach aims to deliver a more engaging and personalized experience, promoting transparency, explanation, and trust~\cite{naveed2020featuristic,faltings2004designing,bostandjiev2012tasteweights,verbert2013visualizing}. Recent efforts in this field leverage various interactive visualizations, employing flow charts~\cite{jin2016go}, Venn diagrams~\cite{parra2014see,andjelkovic2016moodplay}, graph-based representations~\cite{verbert2013visualizing}, cluster maps~\cite{verbert2013visualizing}, concentric circles~\cite{o2008peerchooser}, LineUp-like displays~\cite{gratzl2013lineup,bostandjiev2012tasteweights}, bubbles~\cite{andjelkovic2019moodplay}, maps~\cite{moin2014unified}, and more sophisticated visual analytics systems~\cite{naveed2020featuristic,gretarsson2010smallworlds}. For example, Parra et al.~\cite{parra2014see} empower users to manually blend recommender strategies and adjust their significance, using interactive Venn diagram to inspect the fusion results. TasteWeights~\cite{bostandjiev2012tasteweights}, by combining techniques from traditional recommender system with a novel interactive interface, elucidates the recommendation process and facilitates user preference elicitation. While existing studies concentrate on enhancing end-user experience and trust through process control and flexibility, they often overlook assisting algorithm practitioners in understanding and debugging recommendation algorithms. In contrast, \textit{Prefer2SD} reveals the entire process of multi-stage recommendations (\cref{fig:mult-stage-rec}). It aids algorithm experts in balancing similarity and diversity in friend recommendations, addressing the ``filter bubble''~\cite{pariser2011filter} problem from a different perspective.

\subsection{Human-AI Collaboration for Labeling}
\par Human-AI collaboration in the labeling process~\cite{brachman2022reliance,schaekermann2020human,ashktorab2021ai,zhang2023pathnarratives} is a critical area of research in Human-Computer Interaction (HCI), with approaches like active learning (AL)~\cite{brame2016active} being commonly used. In AL, human annotators label a small subset of data, which is then used to train models and improve performance. Recent advances emphasize the importance of combining both model-based and human-centered perspectives. For instance, Bernard et al.\cite{bernard2018vial} introduced a unified approach that integrates AL with visualization techniques, while Sun et al.\cite{sun2017label} explored ``Label-and-Learn'' interfaces that improve developers' dataset comprehension. Choi et al.~\cite{choi2019aila} took a further step with an interactive attention module that links human labelers with machine learning models through attention-based deep neural networks, enhancing document labeling. These studies highlight the power of combining human expertise with machine learning to create more effective labeling strategies. In addition, tools like KMTLabeler~\cite{wang2024kmtlabeler} and OneLabeler~\cite{zhang2022onelabeler} have been designed to improve the efficiency and accuracy of labeling tasks. KMTLabeler introduces an interactive, knowledge-assisted framework for medical text classification, leveraging human feedback to refine the model while addressing the nuances of medical data labeling. Similarly, OneLabeler incorporates human judgment with machine learning, optimizing the process by assisting annotators with AI-generated suggestions. Both systems demonstrate how interactive human-AI collaboration can make labeling tasks more precise, especially in complex domains like medical text classification. \xiyuan{In our study, human-AI collaboration enables efficient propagation of expert-defined ratios, which is critical for capturing customized player preferences and achieving the desired SD balance in friend recommendations. In addition,} we use a combination of label propagation and active learning~\cite{druck2009active} to propagate expert-defined ratios across a selected group of participants, allowing for efficient human-AI collaboration.

\section{Formative Study}
\subsection{About the Online Service and the Team}
\label{sec: experts}
\par We formed a long-term collaboration with a leading MMORPG development team to explore friend recommendation algorithms. The team includes three male algorithm experts (\textbf{E1-3}), aged $31$, $29$, and $32$, a female game designer (\textbf{E4}), aged $28$, and a female product manager (\textbf{E5}), aged $27$. \textbf{E1}, with five years of experience, focuses on game data analysis and algorithms for friend and gift recommendations. \textbf{E2-3} work on improving recommendation systems. \textbf{E4} creates in-game events and new gameplay, while \textbf{E5} connects the algorithm team with designers. Their goal is to develop more accurate friend recommendation algorithms to strengthen social ties among players.

\subsection{Procedure and Analysis}
\par Over the course of one year, we used participant observation~\cite{blandford2016qualitative} to embed ourselves in the company's routine operations, gaining direct insights into daily workflows and operational bottlenecks. Additionally, we conducted semi-structured interviews with key experts (\cref{sec: experts}), capturing qualitative understanding of their experiences with existing practices.

\par We synthesized insights from these methods using inductive thematic analysis~\cite{braun2012thematic}, which helped us uncover key themes and patterns. Observational data were captured through detailed notes, while interviews, totaling around fifteen hours of audio, were recorded, transcribed via Otter.ai\footnote{\url{http://otter.ai/}}, and manually reviewed for accuracy. Two researchers immersed themselves in the transcripts and observation logs, repeatedly reviewing the data to identify emerging patterns. Each independently performed an initial coding of meaningful segments from the notes and transcripts, allowing the data to guide the process without a predefined framework. They then met to compare and refine their codes, developing a unified coding framework. Potential themes were collaboratively identified and iteratively refined to ensure clarity and distinction. Finally, these findings were synthesized into \textit{conventional practices} and \textit{three key bottlenecks}, which informed the development of five design requirements aimed at addressing these bottlenecks.

\subsection{Findings}
\subsubsection{Conventional Practice}
\par The experts described the conventional method for developing friend recommendation systems, which prioritizes identifying players who are most likely to accept friend requests and build broader social connections within the game. The method typically follows two key stages: \textit{candidate generation} and \textit{ranking}.

\par \textbf{Candidate Generation.} According to \textbf{E1}, the process begins with generating a large pool of potential friend candidates, filtered down from millions of potential matches to a more manageable subset. \textbf{E1} explained, ``\textit{The standard practice is a multi-step process beginning with candidate generation and followed by matching, and finally ranking, which reduces the pool from millions to hundreds and ultimately to tens} (\cref{fig:mult-stage-rec}).'' To further refine the candidate pool, $E2$ and $E3$ emphasized the role of \textit{\textbf{multi-channel matching}} (\cref{fig:mult-stage-rec}(A-B)). This approach integrates various matching criteria to narrow the initial pool from millions to hundreds, leading to a more targeted set of matches.

\par \textbf{Ranking System.} \textbf{E3} explained, ``\textit{After narrowing down the candidate pool, the remaining potential friends are processed through a ranking system} (\cref{fig:mult-stage-rec}(C)).'' This system prioritizes candidates based on metrics like cosine similarity and ranking precision, ultimately reducing the list to fewer than ten recommended friends per user.

\subsubsection{Bottlenecks}
\par Despite the strengths of these conventional methods, the experts highlighted several bottlenecks that hinder the overall effectiveness of friend recommendations.

\par \textbf{B1. Comprehensive Preference Capture.} The current friend recommendation system relies heavily on behavioral data, such as player actions and social interactions, to infer preferences. However, this approach cannot capture the full complexity of player preferences, particularly the more implicit and personalized aspects. Experts suggested that a more holistic approach was needed to understand players' deeper motivations. \textbf{E4} noted, ``\textit{Right now, we're mostly looking at what players do in the game—how they interact, who they connect with—but that only gives us part of the story. There's a whole lot [more] to their preferences that we're not capturing.}'' \textbf{E5} added, ``\textit{Players show their personality in lots of different ways, like picking avatars or customizing their in-game look, but right now, our system doesn't really factor those in.}'' While behavioral data is valuable, it is insufficient on its own. Incorporating additional data, such as symbolic choices like avatars, can improve the accuracy and personalization of recommendations. However, this remains an unaddressed limitation of the current system.

\par \textbf{B2. Similarity Bias.} The current system tends to prioritize players with similar in-game behaviors and interests, which can foster initial connections but also leads to a lack of diversity in social networks. \textbf{E5} observed, ``\textit{The system usually suggests people who are really similar, which works to a certain extent, but it [kind of] limits the range of connections players can make. It's not just about finding someone exactly like you.}'' This bias restricts players' social circles, leading to repetitive recommendations. \textbf{E5} noted, ``\textit{Being constantly advised to associate with the same type of people can lead to a sense of stagnation in a cycle without any new experiences.}'' While fostering connections based on similarity can be beneficial, it also reduces the diversity of interactions, limiting players' exposure to different perspectives. Balancing similarity and diversity in recommendations is a key challenge that the current system has yet to address.

\par \textbf{B3. Group-Specific Needs.} Different player groups—such as new and experienced players—have varying social needs when it comes to friend recommendations. New players often benefit from being matched with similar players to help them build memories of leveling up together, where experienced players seek more diverse connections to explore new content and strategies. The current system lacks the flexibility to adapt to these distinct needs. \textbf{E4} explained, ``\textit{For new [players], having friends with similar interests can help them feel more comfortable. But for veterans, they're often looking for more diverse connections to keep things new.}'' \textbf{E1} added, ``\textit{There's no one-size-fits-all solution. We need a system that can adapt to where players are in their journey and connect them with friends appropriate for that stage.}'' The inability to adjust the recommendations based on player group dynamics is a significant limitation of the existing system.

\subsection{Design Requirements}
\par Building on the conventional practice and identified bottlenecks, we derive five design requirements for enhancing the friend recommendation system.

\par \textbf{DR1. Comprehensive Preference Capture.} This design requirement addresses the challenge of incomplete player preference capture in the current system, which primarily relies on behavioral data (\textbf{B1}). \textbf{E4} highlighted that MMORPGs offer a variety of game experiences, such as \textit{social interactions}, \textit{monster hunting}, and \textit{dungeon challenges}. Avatars also serve as significant forms of self-expression, allowing players to differentiate their characters. Therefore, to improve friend recommendations, the system must capture a wider range of player preferences, going beyond behavioral metrics to include symbolic choices like avatars and other personalized aspects of gameplay.

\begin{figure*}[h]
 \centering 
 \includegraphics[width=\textwidth]{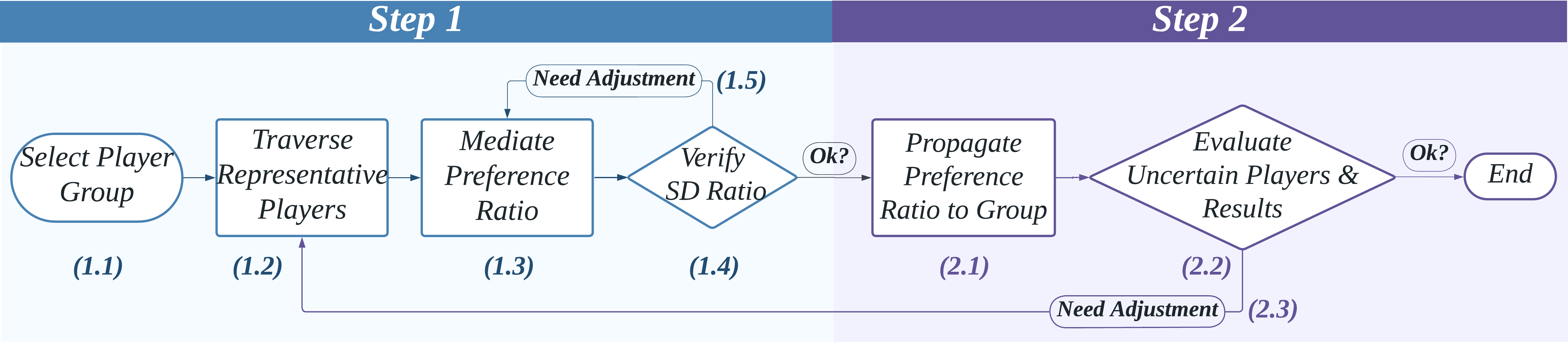}
 \caption[]{Framework of \textit{Prefer2SD}. It introduces a two-step human-in-the-loop workflow. \Stepone{Step 1}: Achieve an optimal SD ratio among representative players in a specific group by mediating their preference ratios. \Steptwo{Step 2}: Extend the preference ratios from \Stepone{Step 1} across the entire player cohort to achieve the desired group-level SD ratio.}
 \label{fig:approach overview}
\end{figure*}

\par \textbf{DR2. Balanced Friend Recommendation Algorithm.} The current bias towards recommending similar friends has led to limited diversity in social connections (\textbf{B2}). Addressing this requires restructuring the algorithm to better balance similarity and diversity. Rather than relying solely on cosine similarity or behavioral overlaps, the recommendation system should incorporate additional factors that promote a wider range of social connections. \textbf{E1} noted that this shift will ensure the diversity of friendships is more reflective of the dynamic game world, fostering richer and more varied interactions among players.

\par \textbf{DR3. Flexible Player Group Selection.} The inability to adapt friend recommendations to the various social needs of distinct player groups—particularly new and experienced players—calls for a more flexible approach (\textbf{B3}). While new players often benefit from connecting with similar players, experienced players typically seek more diverse connections. Therefore, the system should be designed to categorize players flexibly and adjust recommendation parameters based on their current stage and preferences. \textbf{E2} emphasized the importance of this adaptive approach to provide tailored friend recommendations that align with each player's evolving needs.

\par \textbf{DR4. Dynamic Group-Specific Tuning.} The system requires dynamic tuning capabilities to further refine friend recommendations for specific player groups (\textbf{B3}) while addressing the similarity bias (\textbf{B2}). This requirement involves developing an adaptive mechanism that adjusts recommendation parameters in real-time, ensuring the balance between similarity and diversity aligns with the unique preferences of both new and experienced players. As players progress through the game, their social needs evolve, and the system should adapt by offering friend recommendations that reflect these shifts. \textbf{E5} stressed the importance of continuously fine-tuning the balance to sustain player engagement throughout their journey.

\par \textbf{DR5. Continuous Evaluation.} To guarantee the effectiveness, a continuous evaluation mechanism should be established (\textbf{B2}). This would involve ongoing assessment of the system's adjustments, comparing the optimized recommendations against baseline results from traditional approaches. By regularly evaluating the impact of these adjustments, the system can ensure it optimizes for both similarity and diversity, ultimately overcoming the limitations of overly homogeneous recommendations. Experts agreed that this iterative assessment process is critical for maintaining a dynamic and engaging recommendation system for all player types.

\section{Overview of Prefer2SD}
\label{sec: overview}

\par The \textit{Prefer2SD} framework is illustrated in \cref{fig:approach overview}, showing the integration of expert knowledge to mediate preference ratios, aiming to achieve an optimal SD ratio for a specific player group (\textbf{DR2}). The \textit{\textbf{preference ratio}} refers to the proportion of friend candidates coming from different preference channels, representing how friend recommendations are distributed across various preference-based channels. The framework involves a two-step iterative procedure. Specifically, \textbf{\Stepone{Step 1}} first guides experts to select a specific player group (\cref{fig:approach overview}(1.1)) for adjusting the SD ratio, identifying representative players within that group (\cref{fig:approach overview}(1.2)). Subsequently, experts mediate a suitable preference ratio (\cref{fig:approach overview}(1.3)) for each selected representative player to achieve the ideal individual-level SD ratio (\cref{fig:approach overview}(1.4-1.5)). In \textbf{\Steptwo{Step 2}}, \textit{Prefer2SD} propagates the preference ratios established in \textbf{\Stepone{Step 1}} throughout the entire selected player group (\cref{fig:approach overview}(2.1)). Active learning displays a list of players with the most uncertain propagation results (\cref{fig:approach overview}(2.2)), allowing algorithm experts to adjust their preference ratios for further optimization of the group-level SD ratio (\cref{fig:approach overview}(2.3)).
\par The framework is powered by a backend engine and a frontend visualization interface, which are described in the following sections.

\subsection{Backend Engine}
\label{sec: backend}
\par The backend section is structured based on the two steps outlined in \cref{fig:approach overview}. It includes subsections covering \textit{Data Description and Processing}, \textit{Preference Extraction}, \textit{Multi-stage Recommendation}, and \textit{Evaluation Metrics} for \textbf{\Stepone{Step 1}}. Meanwhile, \textbf{\Steptwo{Step 2}} includes the subsections on \textit{Ratio Propagation} and \textit{Active Learning}.

\subsubsection{Data Description and Processing}
\label{sec: data_description}
\par In this study, we analyze a two-month log dataset from an online martial arts-themed MMORPG, spanning from January $1^{st}$ to March $1^{st}$, $2022$, which includes data from $732,437$ players. Monthly updates introduce new social events, gameplay features, and avatar enhancements. Rigorous data masking techniques are employed to ensure user privacy. Daily game logs are extracted based on key characteristics outlined in \cref{sec:mmo}.

\begin{figure*}[h]
 \centering 
 \includegraphics[width=\textwidth]{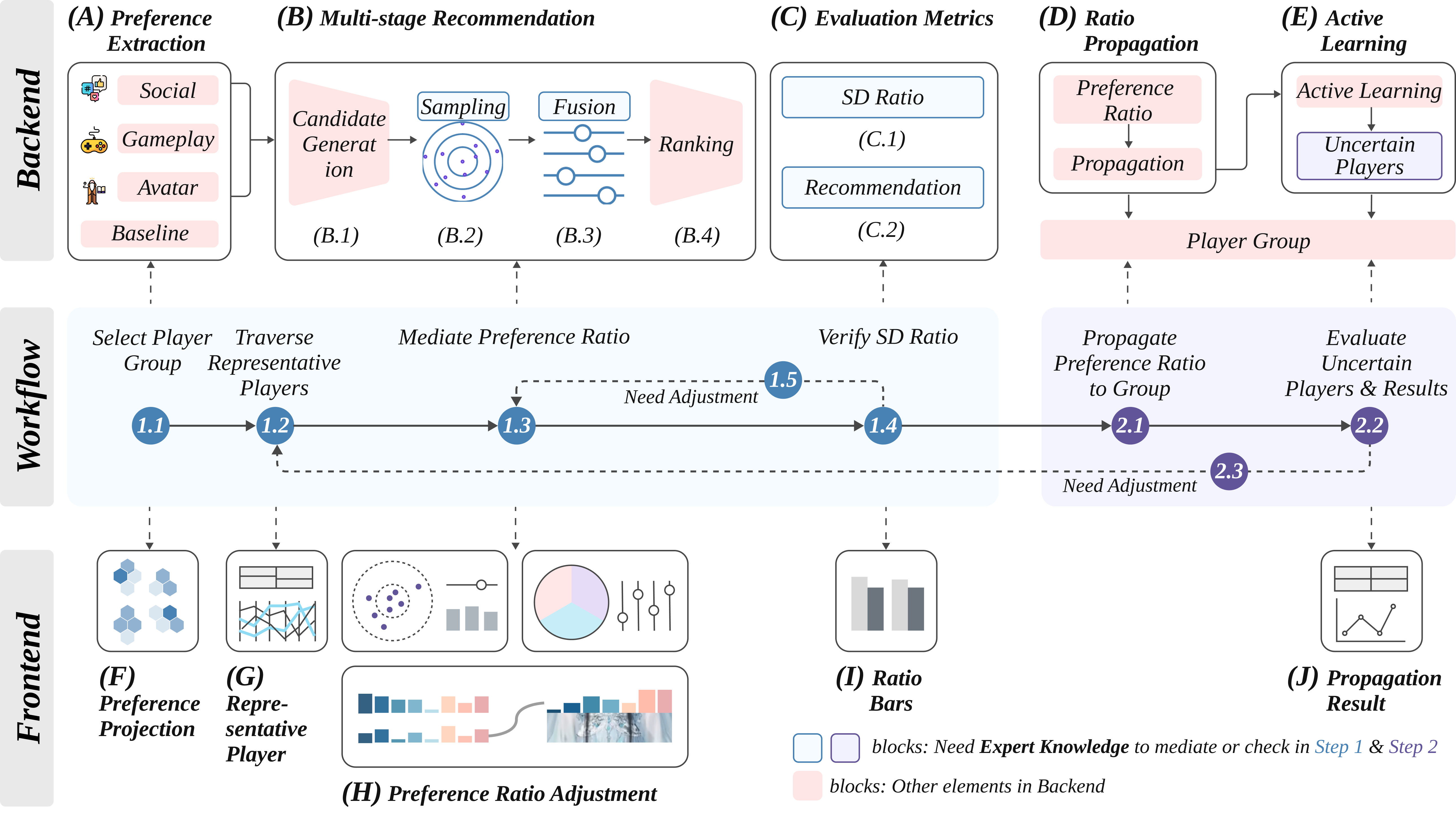}
 \caption[]{Framework of \textit{Prefer2SD} supported by the Backend and the Frontend. (A-C) \& (F-I) Backend and Frontend sections corresponding to \Stepone{Step 1}. (D-E) \& (J) Backend and Frontend sections for \Steptwo{Step 2}.}
 \label{fig:detailed_framework}
\end{figure*}

\subsubsection[]{\Stepone{Step 1} Preference Extraction}
\label{sec: preference_extraction}
\par After consulting with the experts, we derive three types of preference embeddings from the origin game logs: \textbf{social}, \textbf{gameplay}, and \textbf{avatar} (\textbf{DR1}). This approach aims to holistically capture the various facets of players' needs in the game. Furthermore, we obtain a \textbf{baseline} preference derived from their behavioral activities for subsequent comparison (\cref{fig:detailed_framework}(A)).

\par \textbf{Social Preference.} Social networks in MMORPGs reflect players' emotional connections. We extract social preference embeddings from both long-term and short-term behaviors. ``Long-term'' denotes stable social patterns, represented by a cumulative undirected graph of player interactions over a week, generating a $64$-dimensional embedding vector $v_{longterm}$ using \textit{Node2vec}~\cite{grover2016node2vec}. Short-term preferences are captured through daily interaction networks, reflecting transient social behaviors. Given the variability in daily networks, we incorporate three traditional social network metrics: \textit{Page Rank}~\cite{berkhin2005survey}, \textit{KCore}~\cite{dorogovtsev2006k}, and \textit{Closeness Centrality}~\cite{okamoto2008ranking}, assigning a default zero to absent nodes. These metrics form $v_{shortterm}$. The final social preference vector is $P_{social}=CONCAT(v_{longterm},v_{shortterm})$.

\par \textbf{Gameplay Preference.} The variety of gameplay significantly influences MMORPG appeal, including options like \textit{PVE}, \textit{PVP}, and \textit{Guild} (\cref{sec:mmo}). We assess player engagement with each mode on a dichotomous scale, calculating statistical averages over one to seven days to create mean vectors ($v_{1d}$, $v_{3d}$, $v_{5d}$, $v_{7d}$), which are combined into the gameplay preference vector $P_{gameplay} = CONCAT(v_{1d},v_{3d},v_{5d},v_{7d})$. Notably, this method does not rely on user behavioral data, which can be skewed by the game's storyline, leading to similar behavior across players. As preferences can only be expressed at specific points, behavioral measures may not fully capture true player preferences.

\par \textbf{Avatar Preference.} Customizing avatars with various costumes allows players to express personality~\cite{messinger2008relationship}. We analyze game logs for avatar data and visual attire images. From logs, we extract data on owned avatars and acquisition sources, forming three vectors $v_{num}$, $v_{displayed}$, and $v_{acquisition}$. On the visual side, we use a database of $529$ attire images and generate a $768$-dimensional visual embedding $v_{visual}$ with \textit{CLIP}~\cite{radford2021learning}. These vectors support tasks like clustering or image similarity analysis. The final avatar preference vector is $P_{avatar} = CONCAT(v_{num},v_{displayed},v_{acquisition},v_{visual})$.

\par \textbf{Baseline Preference.} Our baseline recommendation system relies on standard game practices for candidate generation (\textbf{DR5}). We extract attributes from players' behavioral records, forming a baseline preference vector for cosine similarity calculations. It allows comparison between original and expert-adjusted recommendation results, highlighting how expert knowledge can improve recommendation diversity and user satisfaction.

\subsubsection[]{\Stepone{Step 1} Multi-stage Recommendation}
\par In our MMORPG recommendation system, we use cosine similarity to generate candidate friends from four preference channels, ensuring a diverse pool (\cref{fig:detailed_framework}(B.1)). We then combine sampling (\cref{fig:detailed_framework}(B.2)) and fusion (\cref{fig:detailed_framework}(B.3)) techniques to balance algorithmic precision with expert insight. Finally, machine learning models (\cref{fig:detailed_framework}(B.4)) are employed to rank the top candidates, ensuring both technical robustness and practical applicability (\textbf{DR2}).

\par \textbf{Candidate Generation.} The four preferences serve as distinct channels during candidate generation. Given the large data volume, we use cosine similarity, computed as $sim_{a,b}=\cos(a,b)=\frac{a\cdot{b}}{|a|\cdot|b|}$, where $a$ and $b$ represent the players’ preference embeddings. The system then identifies potential friends with the highest similarity in each channel.

\par \textbf{Sampling and Fusion.} To enhance recommendation diversity, we integrate expert knowledge to control the ratio of preference channels. \textbf{\textit{1) Sampling: Intra-
Preference Candidate Selection Control.}} It involves organizing candidates in concentric circles based on their cosine similarity to the player~\cite{zhang2006method,10.1145/3565698.3565765}. This structured approach improves diversity beyond random sampling. \textbf{\textit{2) Fusion: Inter-Preference Proportion Control.}} It allows experts to dynamically adjust the proportion of each channel's contribution, ensuring the sum equals one (\cref{fig:detailed_framework}(B.3)). This method blends algorithmic efficiency with human judgment, adapting to varied user needs.

\par \textbf{Ranking.} The ranking phase (\cref{fig:detailed_framework}(B.4)) uses machine learning to produce accurate \textit{top@N} candidates. We train on the first $40$ days of a $60$-day dataset of friend additions, reserving the last $20$ days for testing. We compare two popular ranking algorithms: \textit{GBDT}~\cite{feng2018multi} and \textit{LightGBM}~\cite{ke2017lightgbm}.
For \textit{GBDT}, we set a learning rate of $0.06$, sub-sample rate of $0.56$, and max depth of $9$. \textit{LightGBM} is configured with $31$ leaves and the same max depth. The performance comparison is presented in \cref{tab:table1}. We ultimately select \textit{GBDT} for the ranking phase. 

\begin{table}[h]
	\centering
 \caption{Performance of \textit{GBDT} and \textit{LightGBM}.}
  \vspace{0mm}
    \begin{tabular}{ccccccc} 
\toprule
\textbf{Model} & \textbf{Acc.}   & \textbf{Recall} & \textbf{F1} & \textbf{Precision} &\textbf{AUC}  
\\ 
\midrule
GBDT           & \textbf{76.3\%} & \textbf{0.87}   & \textbf{0.76} & 0.66             & \textbf{0.73}         
\\
LightGBM       & 75.5\%          & 0.83            & 0.75          & \textbf{0.67}              & 0.72         
\\ \bottomrule
\end{tabular}%
 \label{tab:table1}
 \vspace{0mm}
\end{table}

\subsubsection[]{\Stepone{Step 1} Evaluation Metrics}
\label{sec: metrics}
\par We use \textbf{\textit{Diversity}} and \textbf{\textit{Recommendation Quality}} (\cref{fig:detailed_framework}(C)) to assess our friend recommendation system (\textbf{DR5}). \textit{\textbf{Diversity}} (\cref{fig:detailed_framework}(C.1)) is measured using the SD ratio, consisting of \textit{Content Diversity}, \textit{Total\_sim}, and \textit{Fri\_sim}. \textit{Content Diversity}~\cite{carpenter2010study} ensures a wide range of potential connections. \textit{Total\_sim} calculates cosine similarity between candidates and the player, while \textit{Fri\_sim} measures similarity between recommended candidates and the player's friends. Lower values indicate higher diversity. Metrics like \textit{item coverage} and \textit{Gini index} are excluded as they suit item recommendations based on popularity~\cite{klimashevskaia2022mitigating}. \textbf{\textit{Recommendation Quality}} (\cref{fig:detailed_framework}(C.2)) is assessed using \textit{recall}, \textit{precision}, \textit{F1 score}, and \textit{hit rate}. \textit{F1 score} balances precision and recall. \textit{Recall} evaluates the system’s ability to identify relevant friends, while \textit{precision} assesses the relevance. \textit{Hit rate} shows how often a relevant friend is recommended.

\subsubsection[]{\Steptwo{Step 2} Ratio Propagation and Active Learning}
\par Once ratios are assigned to representative players within a group, they are propagated to the entire group using a method similar to label propagation~\cite{iscen2019label}, where the expert-defined preference and SD ratios serve as labels for the group (\cref{fig:detailed_framework}(E)). For samples that are difficult to label automatically, an active learning approach~\cite{druck2009active} is employed (\cref{fig:detailed_framework}(E)). In this step, machine learning algorithms select a subset of difficult-to-label candidates for manual intervention by experts. Experts can then refine the ratios for these specific samples, and perform further iterations of ratio propagation. This human-ai collaboration allows for iterative fine-tuning of the recommendation system, improving the accuracy and adaptability of the recommendations by ensuring that hard-to-label cases are handled with expert input.

\par Once the ratios for the representative players are assigned, these ratios are propagated throughout the player group (\cref{fig:detailed_framework}(D)) using a method akin to label propagation~\cite{iscen2019label}. Label propagation, a semi-supervised machine learning algorithm, generally assigns labels to previously unlabeled data points.
In this context, the predetermined expert-set ratios, i.e, preference ratio and SD ratio (\cref{fig:detailed_framework}(D)), serve as analogous to labels, applied across the player group via label propagation.

\par For samples challenging to label accurately through automated label propagation methods, active learning~\cite{druck2009active} is employed (\cref{fig:detailed_framework}(E)), which filters out the appropriate candidate set for manual labeling through machine learning methods. Expert knowledge is utilized for further secondary ratio setting, and additional iterations of ratio propagation are conducted to achieve targeted and optimal recommendation outcomes. This approach involves human expert intervention alongside machine learning methods, allowing for more precise labeling of these difficult samples. This collaboration enhances the accuracy and efficiency of the system in subsequent iterations.

\subsection{Frontend Visualization}
\par In line with the well-established mantra ``\textit{overview first, zoom and filter, then details on demand}''~\cite{shneiderman2003eyes}, we designed and developed an interactive system to facilitate the exploration of the human-in-the-loop workflow (\cref{fig:approach overview}).

\begin{figure*}[h]
 \centering 
 \includegraphics[width=\textwidth]{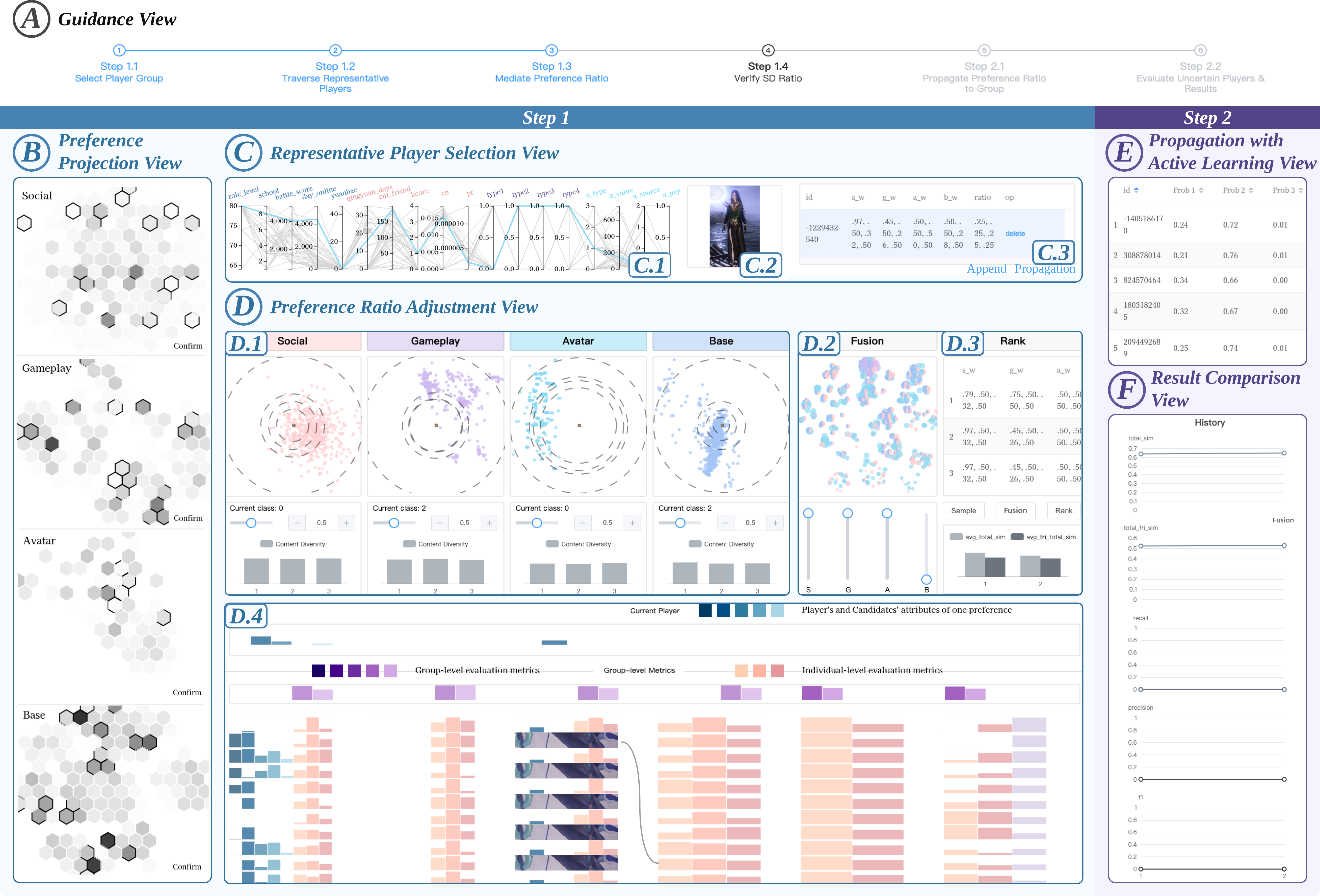}
 \caption[]{The \textit{Prefer2SD} interface incorporates six main views, each serving a distinct purpose. Color-coded labels enhance navigation: blue indicates \Stepone{Step 1}, purple represents \Steptwo{Step 2}, and black highlights guidance. (A) The \textit{Guidance View} provides an overview of progress and directs subsequent actions. (B) The \textit{Preference Projection View} enables selection of specific player groups. (C) The \textit{Representative Player Selection View} assists in choosing a representative player. (D) The \textit{Preference Ratio Adjustment View} supports iterative mediation of preference ratios for ideal SD ratios. (E) The \textit{Propagation with Active Learning View} helps identify uncertain players post-propagation. (F) The \textit{Result Comparison View} summarizes recommendation results across iteration.}
 \label{fig:frontend}
\end{figure*}

\par We provide a \textit{Guidance View} (\cref{fig:frontend}(A)) in the interface, guiding users through each step, excluding adjustments (Steps $1.5$ \& $2.3$). As users progress, active steps turn black, completed steps turn blue, and future steps remain gray. This dynamic navigation helps track users' progress and minimizes confusion.

\par In \textbf{\Stepone{Step 1}}, users begin by navigating through the preference hexbins in the \textit{Preference Projection View} (\cref{fig:frontend}(B)) to identify group-of-interest. Subsequently, in the \textit{Representative Player Selection View} (\cref{fig:frontend}(C)), they analyze specific attributes of the selected group and choose a cohort of representatives for adjusting preference ratios. The \textit{Preference Ratio Adjustment View} (\cref{fig:frontend}(D)) offers two methods-``sampling'' and ``fusion''-for refining intra- and inter-preference candidate selection, which is vital for achieving the desired SD ratio. This process is repeated for each defined player group to ensure precision.

\par \textbf{\Steptwo{Step 2}} focuses on refining ratio propagation through an active-learning-based iterative process. The expert-determined preference ratios are propagated across the remaining players within the same group. Propagation results are reviewed in the \textit{Propagation with Active Learning View} (\cref{fig:frontend}(E)) and \textit{Result Comparison View} (\cref{fig:frontend}(F)), helping identify the most optimal recommendations. Active learning is employed to tackle the challenge of accurately labeling complex, hard-to-replicate players that emerge during \textbf{\Stepone{Step 1}}.

\subsubsection{Preference Projection View}
\par The \textit{Preference Projection View} (\cref{fig:frontend}(B)) offers an overview of the four preferences, aiding experts in selecting player groups (\textbf{DR3}) for targeted intervention in \textbf{\Stepone{Step 1.1}} (\cref{fig:detailed_framework}(F)). Utilizing \textit{t-SNE}~\cite{van2008visualizing}, we project social, gameplay, avatar, and baseline preferences onto a two-dimensional space, where each point corresponds to a player and clusters represent similar preferences. A hexbin visualization overlays the data, with darker colors indicating higher player density. Tooltips display average preferences per hexbin, and hovering highlights related preferences, helping experts quickly identify target groups. Design alternatives are discussed in \cref{hexbin_alter}.

\subsubsection{Representative Player Selection View}
\par The \textit{Representative Player Selection View} (\cref{fig:frontend}(C)) presents detailed information on the four preferences from the previously selected group in the \textit{Projection Hexbin View}, supporting the selection of representative players in \textbf{\Stepone{Step 1.2} }(\cref{fig:detailed_framework}(G)). To effectively handle the high-dimensional preference data, each player's values are displayed through a parallel coordinates plot (\cref{fig:frontend}(C.1)), with attribute names color-coded to correspond to their respective preferences for easy distinction. Experts can narrow down the selection by brushing attribute ranges, which filters the players into a smaller subset displayed in a table (\cref{fig:frontend}(C.3)). Clicking on a row reveals the player's avatar in the center (\cref{fig:frontend}(C.2)), offering a visual reference. Once a player is selected, experts proceed to adjust preference ratios in the \textit{Preference Ratio Adjustment View} to achieve the target SD ratio.

\subsubsection{Preference Ratio Adjustment View}
\par This view (\cref{fig:frontend}(D)) streamlines the multi-stage recommendation process, facilitating experts in iteratively refining preference ratios to achieve the desired SD ratio (\textbf{DR4}) during \textbf{\Stepone{Step 1.3-1.5}} (\cref{fig:detailed_framework}(H-I)). It is divided into the \textbf{upper part} and the \textbf{lower part}. The former integrates three views (\cref{fig:frontend}(D.1-3)) to assist sampling, fusing preferences, and monitoring SD ratio compliance. The latter (\cref{fig:frontend}(D.4)) provides individual-level insights, enabling experts to make informed adjustments to preference ratios based on detailed player data.

\begin{figure}
\centering 
\includegraphics[width=\linewidth]{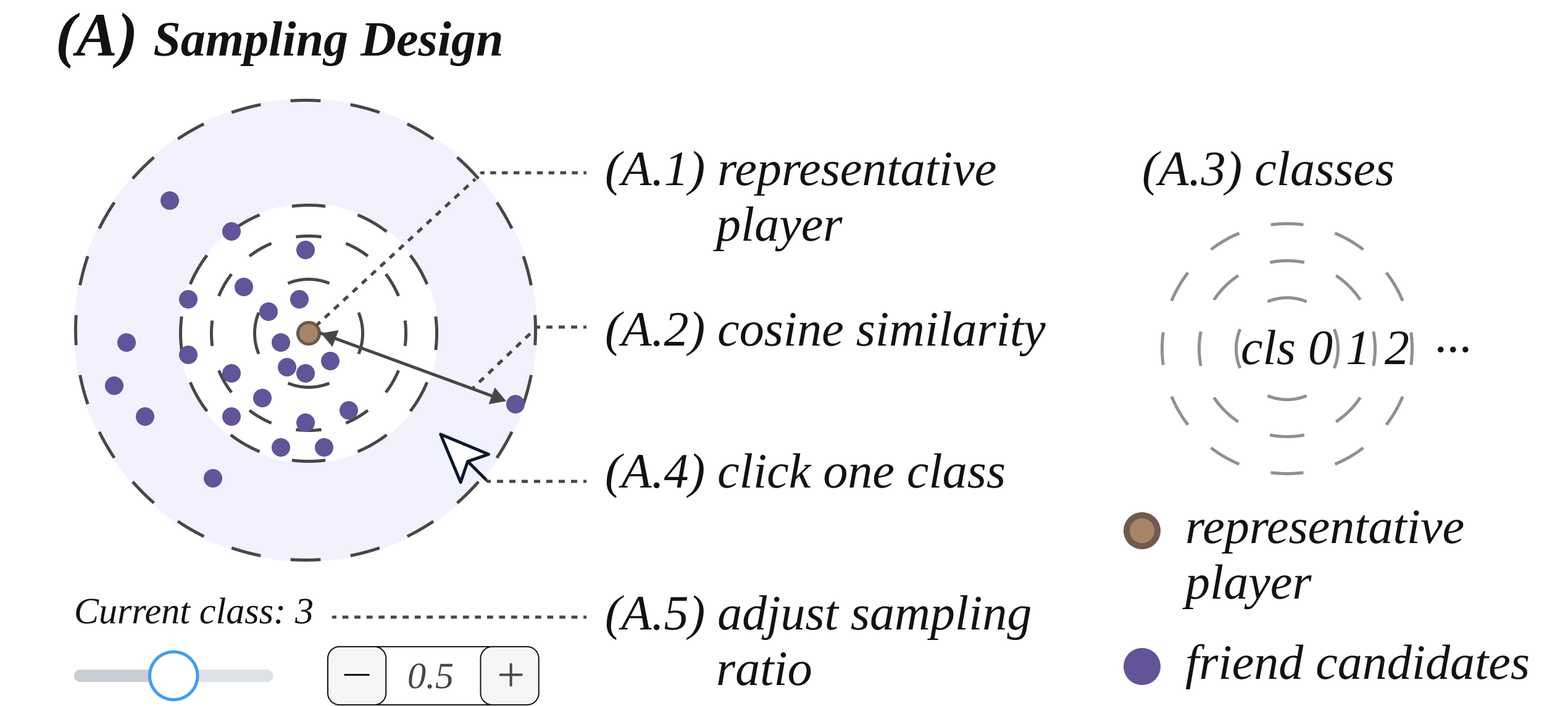}
\caption{Sampling and Fusion Design. (A) Sampling: a scatter plot of candidates created by radial projection. Concentric circles divide the candidates into four classes.}
\label{fig:sampling}
\end{figure}

\par \textbf{Upper Part.} This part is organized into six columns from left to right: the first four columns handle sampling (\cref{fig:frontend}(D.1)), the fifth column (\cref{fig:frontend}(D.2)) focuses on fusion, and the last column is dedicated to evaluating the ranking results (\cref{fig:frontend}(D.3)).

\par \underline{Sampling}. The sampling view (\cref{fig:frontend}(D.1)) is divided into four columns, each representing a different candidate generation channel corresponding to a specific preference. At the top of each column, a radial projection scatter plot visualizes the candidates. Candidates are displayed around a ceteral player (\cref{fig:sampling}(A)), with proximity to the center indicating higher cosine similarity (\cref{fig:sampling}(A.2))). The candidates are arranged in four concentric circles (\cref{fig:sampling}(A.3)), each circle representing a class of candidates. Experts can click on any class (\cref{fig:sampling}(A.4)), to adjust the sampling ratio using a slider bar below (\cref{fig:sampling}(A.5)), enabling precise control over intra-preference sampling. A bar chart beneath the slider provides feedback on content diversity for the selected candidates, aiding experts in tracking the SD ratio as sampling iterates. This approach offers a more refined sampling process compared to random selection, enhancing both precision and diversity.

\par
\begin{figure*}[h]
\centering 
\includegraphics[width=\linewidth]{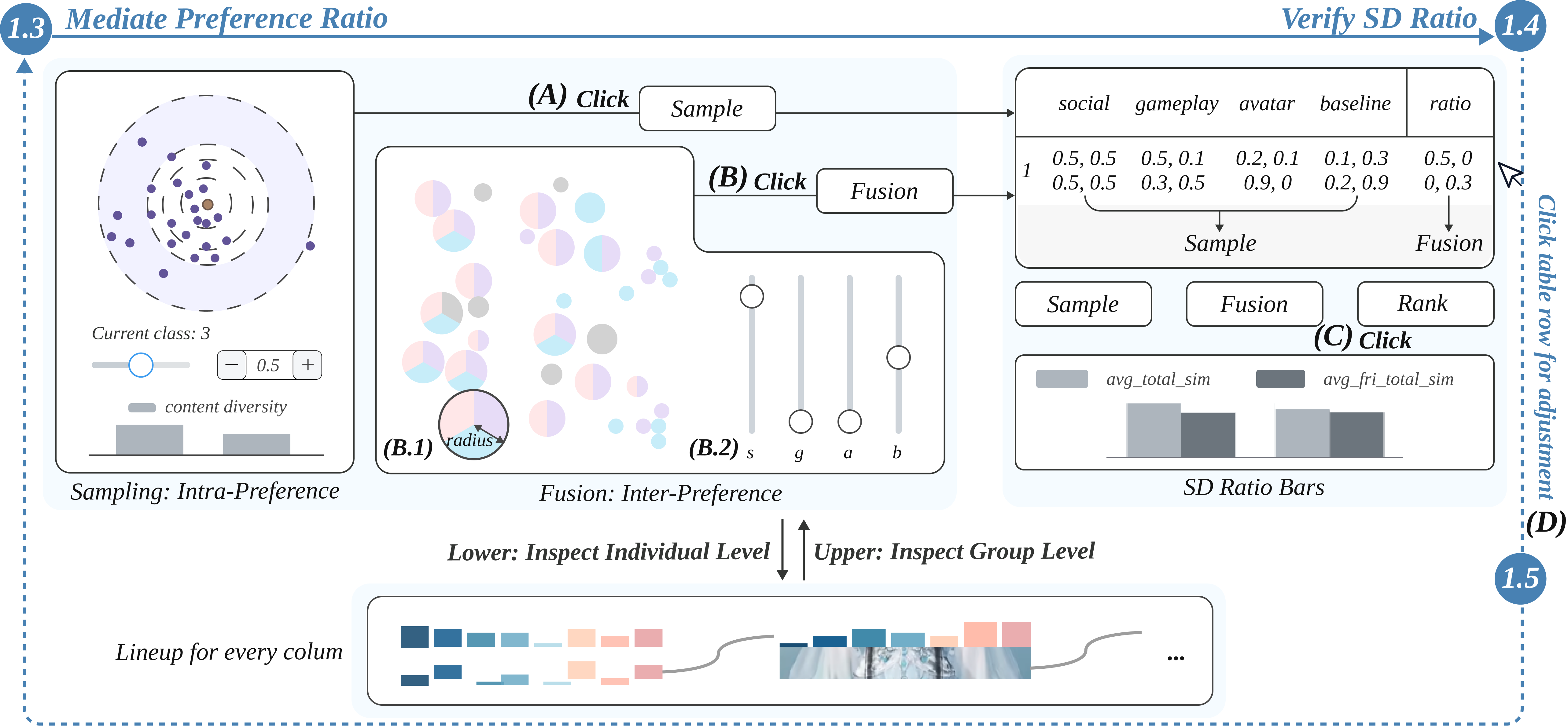}
\caption[]{Interactions in \Stepone{Step 1.3-1.5}: (A-B) Click ``Sample'' and ``Fusion'' button to record intra- and inter- preference ratios to the table. (C) Click ``Rank'' to trigger ranking function to verify SD ratio. (D) Click one table row for preference ratio re-adjustment.}
\label{fig:interaction_columns}
\end{figure*}

 \underline{Fusion.} This view (\cref{fig:frontend}(D.2)) adjusts inter-preference ratios by combining sampled candidates and projecting them onto a $2D$ plane using \textit{t-SNE}~\cite{van2008visualizing}. Similar to the sampling columns, each point represents a candidate; however, in this view, pie charts (\cref{fig:interaction_columns}(B1)) are employed to depict candidates across multiple preferences. Since a player may be selected from various preferences, the pie charts use color coding to match each preference, ensuring a unified and intuitive visual expression. Each pie's radius represents the cosine similarity between the candidate and the central player, with larger radii indicating stronger similarity. This multidimensional visualization aids users in making nuanced inter-preference ratio adjustments, which can be fine-tuned using the four sliders below (\cref{fig:interaction_columns}(B2)). This interactive approach empowers experts to effectively mediate preferences, providing clarity and control.

\par \underline{Ranking and Interactions Among Columns.} The upper section of the Ranking column (\cref{fig:frontend}(D.3)) features a table recording intra- and inter-preference ratios, displaying sampling and fusion results. Users can initiate backend functions for targeted candidate selection by clicking the ``Sample'' and ``Fusion'' buttons (\cref{fig:interaction_columns}(A-B)) located below the table. To finalize the selections, experts can click the ``Rank'' button (\cref{fig:interaction_columns}(C)), which generates a ranked list of recommended friends. A bar chart beneath the table visualizes the SD ratio, calculated using \textit{fri\_sim} and \textit{total\_sim}, providing essential insights for informed decision-making. If a preference ratio leads to an unsatisfying result, experts can click on the corresponding table row for further adjustment (\cref{fig:interaction_columns}(D)). These columns support \textbf{\Stepone{Step 1.3-1.5}}, where the \textbf{upper part} facilitates group-level candidate selection, while the \textbf{lower part} focuses on individual-level adjustments. If modifications are necessary, users can click the corresponding row in the table to alter sampling, fusion, or ranking parameters. The updated SD ratio results are displayed in the bar charts for easy comparison, enabling experts to effectively refine their recommendation outcomes.

\begin{figure}
\centering
\includegraphics[width=\linewidth]{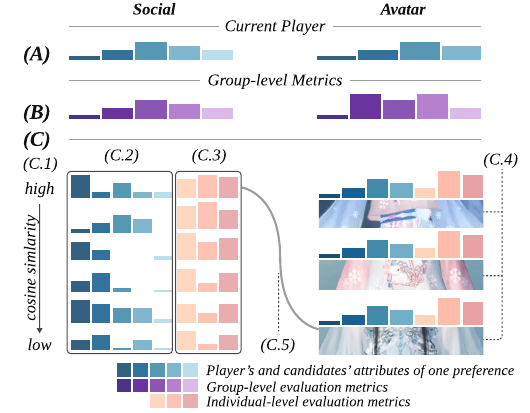}
\caption{Lower part design. (A) Current player attributes. (B) Group-level evaluation metrics. (C) Lineup design showing candidates. (C.1) Row positioning based on cosine similarity. (C.2) Candidates' preference attributes. (C.3) Individual-level evaluation metrics of each candidate. (C.4) Candidates' avatars. (C.5) A line connecting the same candidate appearing in different preferences.}
\label{fig:lineup}
\end{figure}

\par \textbf{Lower Part.} The lower part (\cref{fig:frontend}(D.4)) provides both group- and individual-level recommendation metrics, along with detailed candidate information. This design supports experts in evaluating metrics and individual attributes that have been filtered during the sampling, fusion, and ranking stages (\textbf{DR5}). Following their analysis, experts can modify preference ratios in the upper part. To effectively illustrate multiple attributes, we adopt \textit{Lineup}~\cite{gratzl2013lineup}, organizing each column vertically (\cref{fig:frontend}(D.1-3)), to display detailed candidate information. This arrangement allows users to examine candidate details within each column. An design alternative is discussed in \cref{lineup_alter}.

\par To facilitate comparisons, the top ``Current Player'' section (\cref{fig:lineup}(A)) shows specific preference attributes of the chosen player (\cref{fig:frontend}(C.1)) with an interactive feature to reveal attribute names when hovering over the bars. The middle ``Group-level Metrics'' section (\cref{fig:lineup}(B)) displays average recommendation metrics and SD ratios via bar charts. The bottom \textit{Lineup} design visualizes each candidate's detailed information, with rows representing candidates ranked by cosine similarity (\cref{fig:lineup}(C.1)), and attributes encoded by color within the respective preference columns (\cref{fig:lineup}(C.2)), facilitating direct comparison with the representative. Metrics like individual \textit{cosine similarity}, \textit{total\_sim}, and \textit{fri\_sim} are included to indicate the SD ratio (\cref{fig:lineup}(C.3)). For avatar preferences, they are shown beneath the bar chart (\cref{fig:lineup}(C.4)). In the fusion and ranking columns, only the metrics are shown due to the complexity of the attributes, with predicted outcomes also included. When a candidate appears across multiple columns, the connected curves (\cref{fig:lineup}(C.5)) help compare the rankings across different preferences.

\subsubsection{Propagation with Active Learning View}
\par This view (\cref{fig:frontend}(E)) effectively highlights group players exhibiting the highest levels of uncertainty, encouraging adjustments to the preference ratios following propagation in \textbf{\Steptwo{Step 2.1}}. This organized presentation of uncertainty systematically decreases from top to bottom. In addition, the table displays the probability of each player being assigned to a specific preference ratio, which assists experts in implementing targeted interventions. For interactive analysis, users can select individual players by clicking on corresponding rows. This selection incorporates the chosen players into a dynamic table within the representative player selection interface, alongside previously selected representatives. Consequently, experts can iteratively refine the preference ratios, optimizing the SD ratio more effectively.

\subsubsection{Result Comparison View}
\par The \textit{Result Comparison View} (\cref{fig:frontend}(F)) employs line charts to illustrate the metric results of the entire player group's recommendations following each ratio propagation in \textbf{\Steptwo{Step 2.2}} (\cref{fig:detailed_framework}(J)). The evaluation results are plotted on the $y$-axis, with the $x$-axis indicates the iteration number. Users can hover over the charts to access detailed outcomes for each iteration, facilitating observation and comparison.

\section{Evaluation}
\label{sec: user_study}
\par Upon obtaining institutional IRB approval, we conducted a within-subjects study with $12$ participants to address three research questions aimed at balancing similarity and diversity in friend recommendations.
\par \textbf{RQ1}: How effective and usable is \textit{Prefer2SD} in supporting friend recommendations?
\par \textbf{RQ2}: How does \textit{Prefer2SD} impact users' cognitive load during the friend recommendation process?
\par \textbf{RQ3}: How is the interface design of \textit{Prefer2SD} perceived by users?

\subsection{Participants and Settings}
\par We recruited $12$ experts in friend recommendation algorithms from the collaborating local gaming company to participate in this study, comprising five females and seven males, with ages ranging from $22$ to $40$. Each participant received compensation of $\$10$ per hour. Notably, three participants (\textbf{P1}–\textbf{P3}) were also involved in the formative study. For our baseline, we utilized the conventional method of filtering players through SQL and executing friend recommendations on the company's online platform. Regarding the data settings, we selected data corresponding to February $10^{th}$ from the dataset (\cref{sec: data_description}) to avoid potential data leakage.
On February $10^{th}$, the dataset included $15,339$ players, with an average of $10.28$ friends made in the preceding month and $9.05$ friends made in the following month.

\begin{figure*}[h]
\centering 
\includegraphics[width=\linewidth]{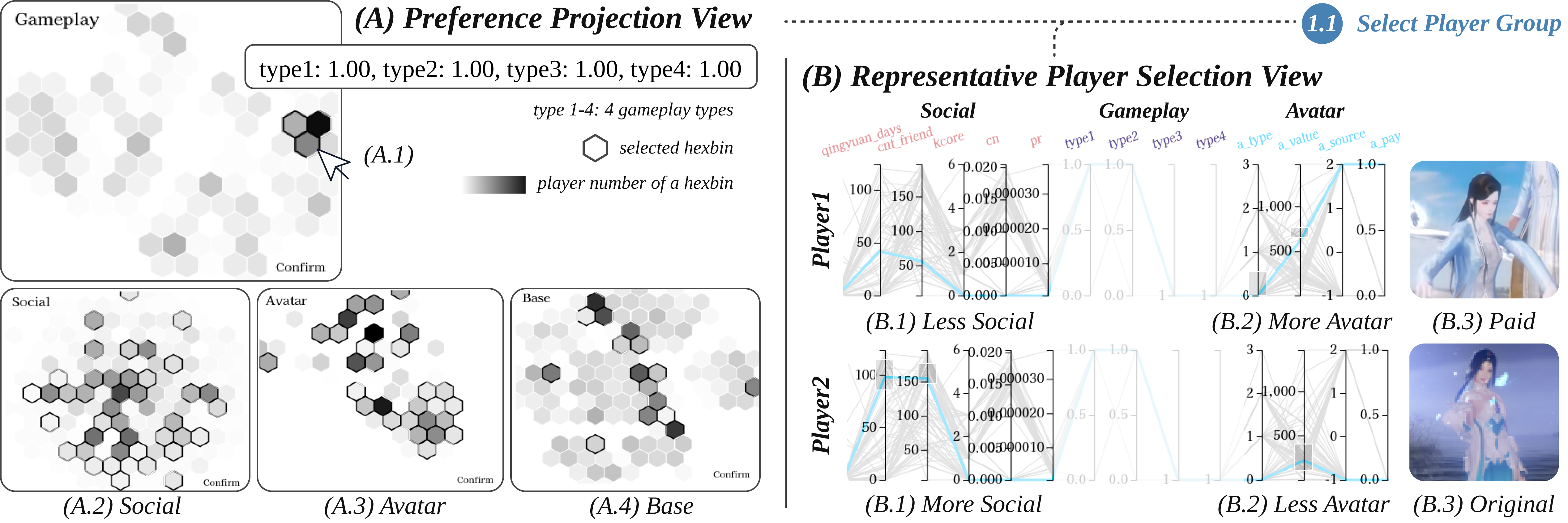}
\caption[]{\Stepone{Step 1.1}: (A) Selection of a cohort with gameplay preferences. (B) Identification of two representative players: one favoring avatars over social aspects, and the other favoring social aspects over avatars.}
\label{fig:case1_1}
\end{figure*}

\subsection{Procedure}
\par One member of our research team conducted a pilot study to assess the completeness and feasibility of the study design. The formal study was structured into four stages. Initially, participants received a brief \textbf{introduction} covering the research background, \textit{Prefer2SD's} workflow, and the interface's visual design and interactions. It included a $5$-minute overview of the research context, followed by a $25$-minute discussion of \textit{Prefer2SD}. Participants then engaged in a phase of \textbf{free exploration}, where they familiarized themselves with \textit{Prefer2SD}, working collaboratively in pairs under a co-discovery learning protocol~\cite{lim1997empirical} for $30$ minutes. During this stage, we observed a notable case that will be elaborated later in \cref{sec: case}. Following exploration, participants completed a \textbf{task} using \textit{Prefer2SD} and the company's platform to recommend suitable friends to players interested in avatar; this task averaged $100$ minutes. Afterward, participants filled out a questionnaire for evaluation. Finally, \textbf{semi-structured interviews} were conducted to gather feedback and suggestions regarding \textit{Prefer2SD}.

\subsection{Measurement}
\par We employed a $7$-point Likert scale ($1$: strongly disagree, $7$: strongly agree) to assess \textit{Prefer2SD}'s effectiveness, usability, cognitive load, and interface design. For assessing \textbf{effectiveness and usability}, we adapted several items from the \textit{System Usability Scale}~\cite{bangor2009determining}, including integration of functions, ease of learning, ease of use, willingness to use again, and confidence in using the system. Additionally, we measured the system's ability to balance similarity and diversity, its efficiency in generating recommendations, and participants' willingness to recommend it to others. To measure \textbf{cognitive load}, we utilized the \textit{NASA-TLX questionnaire}~\cite{hart2006nasa}, which evaluates cognitive demands across six dimensions: mental demand, physical demand, temporal demand, performance, effort, and frustration.
For the \textbf{interface design}, we evaluated both the visual the interaction aspects of \textit{Prefer2SD}.

\par During task completion, we recorded participants' actions via screen and audio capture while they engaged in a think-aloud protocol~\cite{jaaskelainen2010think}, verbalizing their thoughts and reasoning. We also monitored \textbf{task completion time} for both the baseline and \textit{Prefer2SD}, as well as \textbf{recommendation quality} and \textbf{diversity}. Task quality was measured using several metrics, including overall similarity with ground truth (GT) friends, which indicated how closely the recommended friends matched those that participants actually added in the subsequent month. In terms of diversity, we considered the total similarity between the player and the recommended friends (\textit{total$\_$sim}), as well as the total similarity between the players' friends and the recommended friends (\textit{total$\_$fri$\_$sim}). This approach prioritized the long-term social connections between users and their recommended friends, rather than relying solely on traditional metrics like \textit{hit rate} or \textit{recall}, which primarily focus on short-term interactions.

\subsection{Case Study: Suggesting a Varied Set of Friends}
\label{sec: case}
\par This case study involved collaboration with \textbf{P1} and \textbf{P9}, focused on using \textit{Prefer2SD} to suggest diverse friends for core players to enhance their satisfaction and sense of belonging.

\par \textbf{Goal: Enhancing Social Connectivity for Core Players.} The goal was to address the needs of core players, a group with strong gameplay preferences. However, recommending friends to this group was challenging due to the limited pool of skilled players. Continuous recommendations of similar individuals could create a ``filter bubble''~\cite{pariser2011filter}, where social interactions are limited to those with comparable skills and attributes. To solve it, \textbf{P9} proposed promoting diverse acquaintances to enrich their social experience.

\begin{figure*}[h]
\centering 
\includegraphics[width=\linewidth]{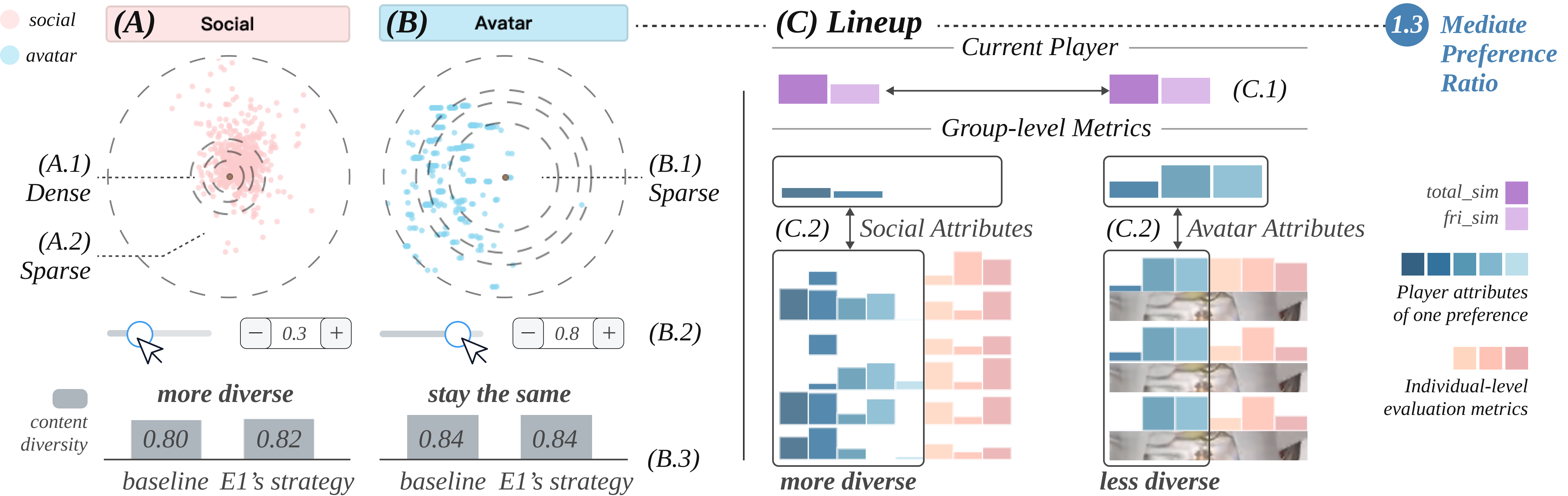}
\caption[]{\Stepone{Step 1.3} (A-B) Conducting Sampling
for both social and avatar preferences and evaluating metric values. (C) Delving deeper into candidate attributes from both preferences and comparing metrics at both group and individual levels.}
\label{fig:case1_2}
\end{figure*}

\par \textbf{\Stepone{Step 1.1}} \textbf{Exploring a Cohort with Gameplay Preference.} 
\textbf{P1} and \textbf{P9} imported three months of data (\cref{sec: data_description}), focusing on a weekend period. \textbf{P1} explained, ``\textit{Weekends usually have more players active, that's why we're focusing on them for our study.}'' They then directed to the hexbins in the \textit{Preference Projection View} (\cref{fig:case1_1}(A)), quickly identifying a cluster of skilled players with gameplay type values near $1$, spanning three distinct hexbins(\cref{fig:case1_1}(A.1)). 
Further analysis revealed that these players had diverse preferences in social, avatar, and base attributes, as highlighted in each corresponding view (\cref{fig:case1_1}(A.2-4)). 
\textbf{P9} remarked, ``\textit{It's completely natural since everyone has their own preferences.}'' 
They then integrated the players in the selected hexbins into the \textit{Representative Player Selection View} (\cref{fig:case1_1}(B)) for further analysis. 
Using the parallel coordinate plot, \textbf{P1} and \textbf{P9} evaluated the group across different preference attributes, selecting two representative players—one favoring avatar over social, and the other favoring social over avatar (\cref{fig:case1_1}(B.1-2)). Additionally, they noted that the avatar-preferring player had chosen a paid version, while the social-preferring player opted for the original version (\cref{fig:case1_1}(B.3)), reinforcing their earlier observations.

\par \textbf{\Stepone{Step 1.2}} \textbf{Examining two representative players.} \textbf{P1} and \textbf{P9} began to examine two chosen representative players and chose the player with the avatar preference first. To increase recommendation diversity, they decided to recommend candidates from two preferences, \textbf{social} and \textbf{avatar}.

\par \textbf{\Stepone{Step 1.3}} \textbf{Mediating intra-preference ratio: Sampling.} 
In the \textit{Preference Ratio Adjustment View}, they first explored the sampling columns (\cref{fig:case1_2}). \textbf{P1} observed that social preference candidates were tightly clustered within the inner three circles (\cref{fig:case1_2}(A.1)), indicating higher similarity, while the outermost circle showed more diversity due to the sparsity of the candidates (\cref{fig:case1_2}(A.2)).
To standardize the process, they initially set the sampling frequency for each class to $1$, feeding all candidates to ranking and click the ``Sample'' button.
They then adjusted the frequency of each concentric circle, reducing it for the inner circles and increasing it for the sparse outer circle (\cref{fig:case1_2}(A.3)). Specifically, the intra-preference ratio for \textbf{social} was adjusted to $0.3$, $0.3$, $0.3$ and $0.8$ from inner to outer classes, they then evaluated the results.
Content diversity improved slightly, from from $0.80$ to $0.82$.
For the \textbf{avatar} preference, they started with a baseline sampling frequency of $1$ (\cref{fig:case1_2}(B.2)) and noted that the points were more spread out compared to social (\cref{fig:case1_2}(B.1)).
Thus, they increased the sampling frequency to $0.8$ for each class but found no significant change in \textit{content diversity} (\cref{fig:case1_2}(B.3)), suggesting that adjusting the candidate sampling frequency had minimal impact.

\par To delve deeper into the candidates distribution for the two strategies, \textbf{P1} and \textbf{P9} shifted to the lineup view below (\cref{fig:case1_2}(C)). At the group level, the \textit{total\_sim} values were nearly identical, yet social candidates showed broader diversity in \textit{friend\_sim} (\cref{fig:case1_2}(C.1)). Upon reviewing individual attributes, they confirmed that social candidates exhibited more diversity compared to avatar ones (\cref{fig:case1_2}(C.2)), consistent with their earlier findings during sampling (\cref{fig:case1_2}(A-B)).

\begin{figure}
\centering
\includegraphics[width=\linewidth]{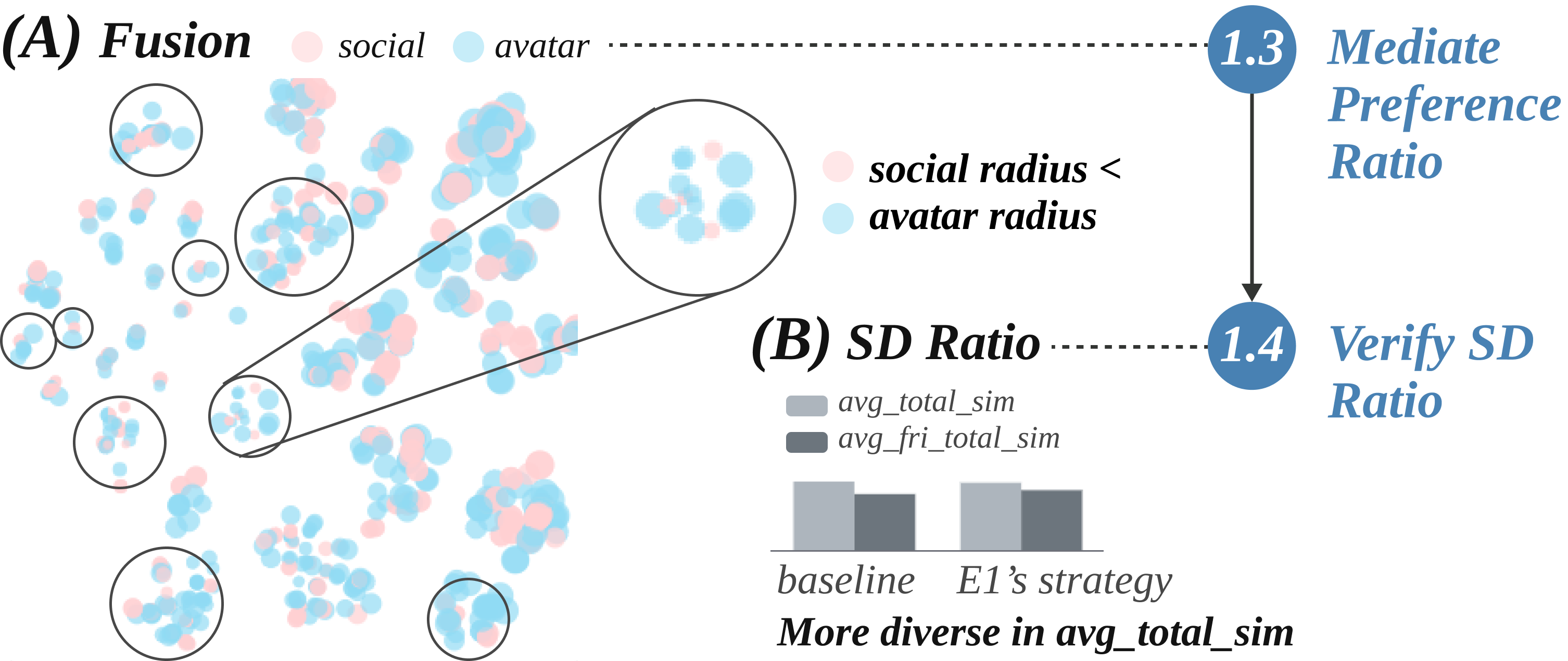}
\caption[]{\Stepone{Step 1.3-4} (A) Exploring pies of social and avatar candidates in Fusion. (B) Comparing SD ratio between results of baseline and experts' own strategy.}
\label{fig:case1_3}
\end{figure}

\par \textbf{\Stepone{Step 1.3}} \textbf{Mediating inter-preference ratio: Fusion.} 
During fusion, the experts analyzed the projected players in the fusion column. They noted that while social and avatar candidates had nearly identical distributions, social candidates had smaller radii overall.
In \cref{fig:case1_3}(A), two key observations merged: 1) Few candidates were selected from both preferences, as most pies were dominated by a single color; 2) Social candidates had smaller radii than avatar candidates, suggesting greater diversity. Based on these findings, they adjusted the fusion ratio to prioritize social candidates, setting it at $0.7$ for social and $0.3$ for avatar to ensure a more diverse candidate pool.

\par
\begin{figure*}[h]
\centering 
\includegraphics[width=\linewidth]{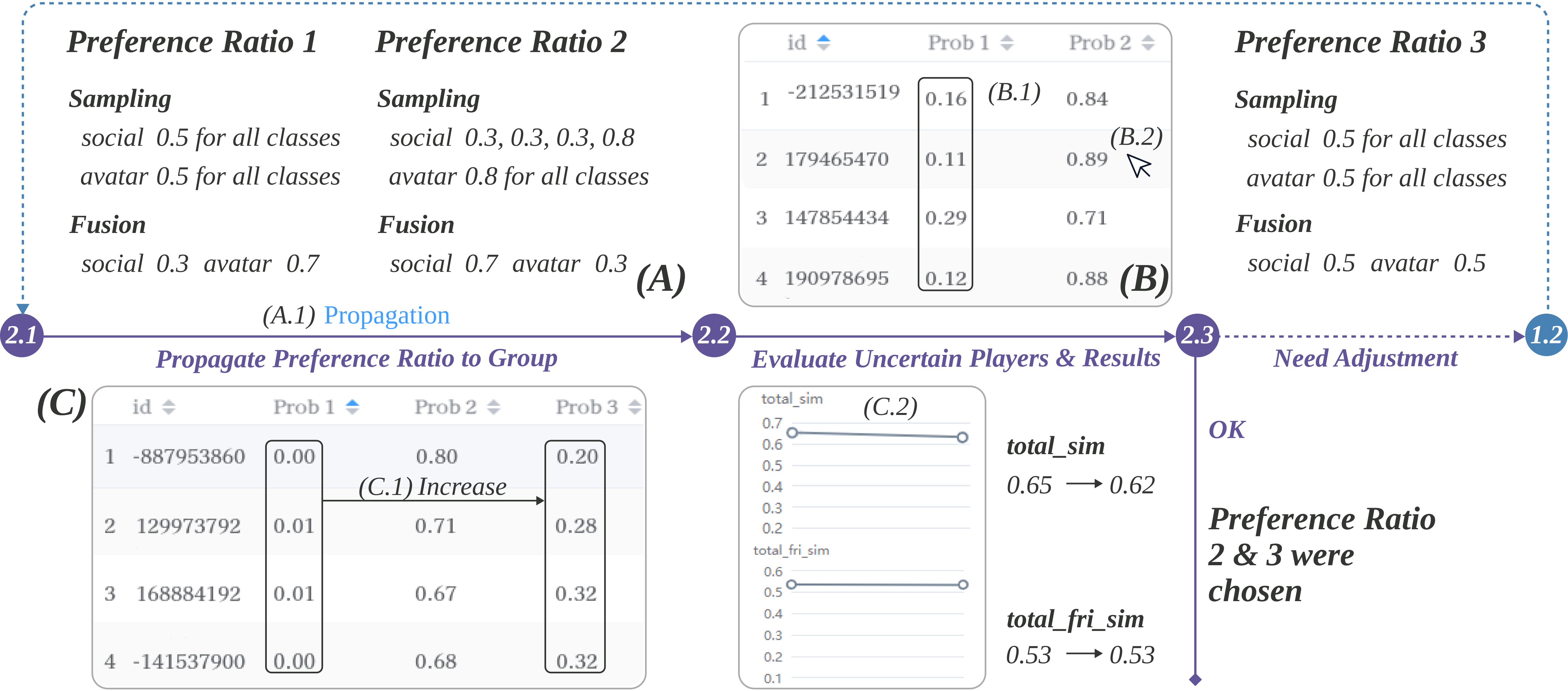}
\caption[]{\Steptwo{Step 2.1-3} (A) Propagate \CaseOneRatio{Preference Ratios $1$ \& $2$} to the player group. (B) Identify the most uncertain player for adjusting the preference ratio and propagate again. (C) Adopt \CaseOneRatio{Preference Ratios $2$ \& $3$} based on the favorable probabilities and metric values.}
\label{fig:case1_4}
\end{figure*}

 \textbf{\Stepone{Step 1.4}} \textbf{Verify SD ratio.}
After fusion, the results were fed into the ranking model to assess the SD ratio. For comparison, they used the baseline preference incorporated, representing their conventional method. The intra- and inter-preference ratios for baseline candidates were set to $1$, and all others to $0$, ensuring only baseline candidates entered the ranking stage. The results showed that their strategy achieved a lower \textit{total\_sim} compared to the baseline, while \textit{fri\_sim} remained unchanged (\cref{fig:case1_3}(B)). This indicated a more diverse recommendation. Consequently, they assigned the intra- and inter-preference ratio as \CaseOneRatio{Preference Ratio $1$} for this player (\cref{fig:case1_4}(A)).

\par \textbf{\Stepone{Back to Step 1.2}} \textbf{Traverse to the second representative.}
\textbf{P1} and \textbf{P9} applied the iterative mediation process to the second representative, who favored social over avatar preferences. Through experimentation, they found that setting the intra-preference ratios of both social and avatar classes to $0.5$, and the inter-preference ratios of social and avatar to $0.3$ and $0.7$, respectively, resulted in a more diverse set of friend recommendations. Consequently, they identified this ratio configuration as \CaseOneRatio{Preference Ratio $2$} (\cref{fig:case1_4}(A)). At this stage, the experts had successfully assigned two preference ratios to the selected player cohort based on their expertise.

\par \textbf{\Steptwo{Step 2.1-2}} \textbf{Preference Ratio Propagation and Evaluation.} After assigning two preference ratios, the experts activated the \textit{Propagation} button (\cref{fig:case1_4}(A.1)) to apply these ratios across the player cohort. Players flagged through active learning for re-mediation were displayed in the \textit{Propagation with Active Learning View} (\cref{fig:case1_4}(B)). \textbf{P9} observed that the probability of propagating the \CaseOneRatio{Preference Ratio $1$} was extremely low, with most values under $0.2$ (\cref{fig:case1_4}(B.1)). As a result, they concluded that this ratio should be considered an outlier and unsuitable for propagation.

\par \textbf{\Steptwo{Step 2.3}} \textbf{Iterative Adjustment.}
The experts then selected the player with lowest possibility ($0.11$ for \CaseOneRatio{Preference Ratio $1$}) for re-mediation (\cref{fig:case1_4}(B.2)). After observations in the relevant views from \textbf{\Stepone{Step 1.3-5}}, they assigned this player an intra-preference ratio of $0.5$ for all classes, and an inter-preference ratio of $0.5$ for both social and avatar classes, designating it as \CaseOneRatio{Preference Ratio $3$}. Upon executing propagation again (\cref{fig:case1_4}(B.2)), they observed a significant increase in propagation probability (\cref{fig:case1_4}(C.1)).
The recommended outcomes were more diverse (\cref{fig:case1_4}(C.2)), with \textit{total\_sim} decreased from $0.65$ to $0.62$ while \textit{total\_fri\_sim} remained constant. Based on the iterative process, \textbf{P1} and \textbf{P9} concluded that the \CaseOneRatio{Preference Ratio $2$ \& $3$} were appropriate for the player group.

\par \textbf{Summary.} The experts recognized the value of \textit{Prefer2SD} for friend recommendations, noting that visualizing candidate information at different granularities made analysis and adjustments more intuitive, thereby enhancing their efficiency.

\section{Results and Analysis}
\par In our quantitative analysis, we conducted the Wilcoxon signed-rank test~\cite{oyeka2012modified} to evaluate the differences in participants' ratings across various factors between the two systems. For the qualitative analysis, we utilized inductive thematic analysis~\cite{braun2012thematic} on the interview transcripts. Two researchers independently coded the data, followed by a cross-checking of themes to ensure consistency and reliability. The findings were reported in relation to the three research questions.

\begin{table*}[h]
    \caption{The statistical feedback with \textit{Baseline} and \textit{Prefer2SD}, where the $p$-values (n.s: not significant, *: $p<.050$, **: $p<.010$, ***: $p<.001$) is reported.}
    \centering
    \begin{tabular}{llccccc}
        \toprule
        \multirow{2}{*}{\textbf{Category}} & \multirow{2}{*}{\textbf{Factor}} & \small{\textbf{Baseline}} &\small{\textbf{Prefer2SD}} & \multirow{2}{*}{$W$} & \multirow{2}{*}{$p$} & \multirow{2}{*}{$Eff. Size$}\\
        &  & \small{\textbf{Mean/S.D.}} &\small{\textbf{Mean/S.D.}} & &\\
        \midrule
        \multirow{3}{2cm}{Effectiveness} 
        & Balancing similarity and diversity & $2.00/0.95$ & $5.91/0.99$ & $0.00$ & $0.002$** & $0.88$\\
        & Improving efficiency & $3.25/1.22$ & $5.33/1.23$ & $0.00$ & $0.008$** & $0.88$\\
        & Functions & $4.17/1.40$ & $5.83/1.02$ & $7.00$ & $0.020$* & $0.69$\\
        \midrule
        \multirow{4}{2cm}{Usability} 
        & Easy to learn & $4.25/1.14$ & $3.41/1.37$ & $15.50$ & $0.124$n.s & $0.46$\\
        & Easy to use & $3.33/1.55$ & $4.42/0.99$ & $14.00$ & $0.174$n.s & $0.42$\\
        & Willing to recommend & $3.25/1.22$ & $5.50/1.31$ & $2.00$ & $0.006$** & $0.82$\\
        & Willing to use again & $3.33/1.61$ & $5.67/1.23$ & $0.00$ & $0.003$** & $0.88$\\
        & Confident to use & $4.17/1.47$ & $6.00/0.95$ & $4.50$ & $0.011$* & $0.75$\\
        \midrule
        \multirow{6}{2cm}{Cognitive Load}
        & Less Mental & $3.17/1.34$ & $5.00/1.47$ & $1.50$ & $0.014$* & $0.81$\\
        & Less Physical & $4.17/0.94$ & $5.58/1.08$ & $4.00$ & $0.009$** & $0.79$\\
        & Less Temporal & $3.91/1.31$ & $4.92/1.08$ & $20.00$ & $0.132$n.s & $0.43$\\
        & Performance & $2.42/1.08$ & $5.50/1.17$ & $0.00$ & $0.003$** & $0.88$\\
        & Less Effort & $4.00/1.21$ & $5.17/1.27$ & $6.00$ & $0.020$* & $0.68$\\
        & Less Frustration & $3.33/1.49$ & $4.75/1.54$ & $26.00$ & $0.060$* & $0.55$\\
        \midrule
        \multirow{2}{2cm}{Design}
        & Visual design & - & $6.10/0.86$ & - & - & -\\
        & Interaction design & - & $6.28/0.63$ & - & - & -\\
        \bottomrule
    \end{tabular}
    \label{tab:user study}
\end{table*}

\subsection{RQ1: How effective and usable is \textit{Prefer2SD} in supporting friend recommendations?}
\par \textbf{Effectiveness.} \textit{Prefer2SD} showed significant improvements in balancing similarity and diversity as well as improving recommendation efficiency ($W=0.00$, $p<0.01$ for both). Additionally, participants perceived that the system's functions were well integrated into the interface ($W=7.00$, $p<0.05$) compared to the baseline system in \cref{tab:user study}. Most participants ($11/12$) found the framework of \textit{Prefer2SD} very familiar and easy to grasp. \textbf{P12} expressed confidence that introducing mediation to intra- and inter-preference ratios in \Stepone{Step 1} would effectively control the SD ratio, a crucial factor in achieving more balanced recommendation outcomes. ``\textit{But it requires a deep understanding of each player's personal desires.}'' remarked \textbf{P9}. Additionally, \textbf{P2} emphasized the significance of \Steptwo{Step 2}, stating, ``\textit{It allows us to refine recommendations, enhancing the system's accuracy and flexibility.}''

\par
\begin{figure*}[h]
\centering 
\includegraphics[width=\linewidth]{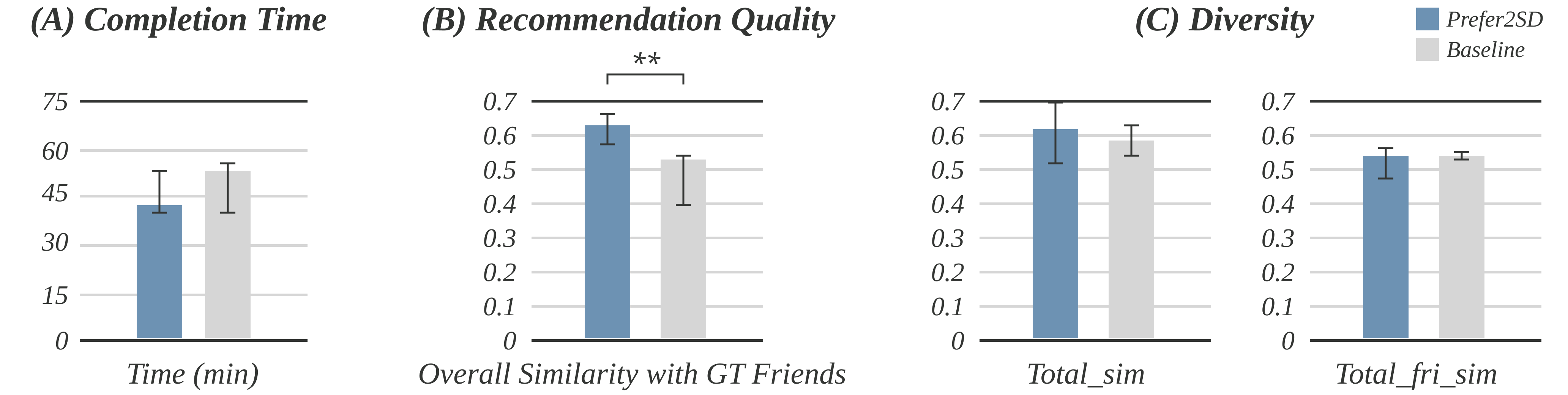}
\caption[]{\textit{Prefer2SD} outperformed the baseline in three dimensions: (A) Completion time; (B) Recommendation quality; (C) Diversity. (*: $p<.050$, **: $p<.010$, ***: $p<.001$)}
\label{fig:task}
\end{figure*}

 As shown in \cref{fig:task}, the task completion time was slightly reduced when using \textit{Prefer2SD} compared to the baseline, indicating some improvement in efficiency (\cref{fig:task}(A)). More importantly, recommendation quality significantly increased (\cref{fig:task}(B)), with a higher match between recommended friends and actual friends added by the player ($W=0.00$, $p<0.01$). This suggests that \textit{Prefer2SD} can provide more accurate and sustainable friend recommendations, which contribute to the development of long-term social networks. Several participants acknowledged these improvements. \textbf{P3} mentioned, ``\textit{Prefer2SD's ability to balance similarity and diversity in the recommendations is very noticeable. It's an improvement compared to typical systems that usually lean too much towards similarity.}'' \textbf{P7} added, ``\textit{It is often difficult to achieve in our conventional practice.}'' 

\par As for diversity (\cref{fig:task}(C)), the results between the two systems were relatively similar. This indicated that the player group who strongly prefer avatar tended to choose friends who were more similar to themselves. Several participants offered their insights. \textbf{P6} pointed out, ``\textit{It's important to recognize that while diversity is a key goal, too much diversity without sufficient similarity might lead to user disengagement. The system must find a balance that satisfies both diversity and personal relevance}''.

\par \textbf{Usability.} The usability analysis revealed notable improvements with \textit{Prefer2SD} over the baseline system across several key areas. Although no significant differences were observed in terms of ease of learning ($W=15.50$, $p>0.05$) and ease of use ($W=14.00$, $p>0.05$), participants became increasingly adept as they familiarized themselves with \textit{Prefer2SD}'s interface. While the initial design presented some challenges, participants adapted over time. For instance, \textbf{P3} highlighted that the \textit{Guidance View} (\cref{fig:frontend}(A)) played a pivotal role in this adaptation, connecting the familiar framework with each view and enhancing overall system comprehension.

\par \textit{Prefer2SD} demonstrated clear strengths in usability metrics. Participants were significantly more likely to recommend the system to others ($W=2.00$, $p<0.01$) and showed a notably higher willingness to use it again ($W=0.00$, $p<0.01$), indicating a strong preference for \textit{Prefer2SD}. Many participants ($8/12$) praised the system’s versatility, highlighting its adaptability to other game friend recommendation scenarios. For instance, \textbf{P9} noted that \textit{Prefer2SD}’s flexible design could seamlessly integrate into different game environments, making it highly adaptable across various platforms. Additionally, they reported a significant boost in confidence when using \textit{Prefer2SD} ($W=4.50$, $p<0.05$) compared to the baseline system. All participants ($12/12$) pointed out that the system's diversity indicators facilitated a more intuitive assessment of the variety in recommendations, simplifying the evaluation of its outputs.

\par These findings indicate that although both systems exhibited comparable learning curves and ease of use, \textit{Prefer2SD} outperformed in fostering user satisfaction and engagement. Participants were more inclined to recommend and reuse the system, demonstrating increased confidence in its application.

\subsection{RQ2: How does \textit{Prefer2SD} impact users' cognitive load during the friend recommendation process?}
\par \textit{Prefer2SD} significantly reduced cognitive load across multiple dimensions compared to the baseline system. Participants reported experiencing less mental ($W=1.50$, $p<0.05$) and physical load ($W=4.00$, $p<0.01$) when using \textit{Prefer2SD} , indicating that the system required less cognitive processing for decision-making during the friend recommendation process. \textbf{P5} specifically noted that, compared to the company's node-based recommendation platform, \textit{Prefer2SD}'s visual interface provided a clearer understanding of the recommendation process. ``\textit{Having a visual representation of the entire recommendation flow makes it much easier to grasp what's happening at each step,}'' said \textbf{P5}.

\par The results also showed an improvement in temporal load, although the difference between \textit{Prefer2SD} and the baseline system was not statistically significant ($W=20.00$, $p>0.05$). \textbf{P1} explained, ``\textit{The system provides more detailed information and gives me more control over the recommendation process, but it does take a bit longer to explore and fully understand everything.}'' Participants, however, reported a significant improvement in performance ($W=0.00$, $p<0.01$), suggesting that they felt more effective and successful in completing tasks using \textit{Prefer2SD}. As \textbf{P11} noted, ``\textit{With Prefer2SD, I can dynamically adjust the granularity of the recommendations and tailor strategies for different player groups. This level of control is invaluable for reducing player churn and enhancing the core user experience.}'' \textit{Prefer2SD} also required significantly less effort from users ($W=6.00$, $p<0.05$), further reducing cognitive strain during the friend recommendation process.

\par Lastly, \textit{Prefer2SD} led to a significant reduction in frustration levels compared to the baseline system ($W=26.00$, $p<0.05$). Several participants appreciated how \textit{Prefer2SD}'s visual tools enabled them to conduct in-depth analyses without the need to switch between multiple platforms. ``\textit{The ability to perform deeper analysis within a single interface saves us from jumping between different systems, which often causes confusion and delays,}'' remarked \textbf{P4}. This integration streamlined their workflow, helping them avoid the frustration of managing multiple tools during the analysis process.

\par In summary, \textit{Prefer2SD} significantly reduced participants' overall cognitive load, particularly in terms of mental, physical, and effort-related dimensions. By improving performance while simultaneously lowering effort and frustration, the system effectively streamlined the friend recommendation process, making the task more manageable and less cognitively taxing for users.

\subsection{RQ3: How is the interface design of \textit{Prefer2SD} perceived by users?}
\par The interface design of \textit{Prefer2SD} received highly positive feedback from participants, as reflected in the strong ratings for both visual and interaction design. The visual design achieved an impressive mean score of $6.10$ ($SD=0.86$), while the interaction design received an even higher rating of $6.28$ ($SD=0.63$), indicating that participants found the system not only visually appealing but also highly intuitive and user-friendly.

\par A key aspect of the interface that resonated with participants was the integration of hexbins and tooltips within the \textit{Preference Projection View}. \textbf{P1} particularly praised this feature, emphasizing its effectiveness in quickly identifying player groups that required specialized intervention. This visual tool allowed experts to efficiently navigate large datasets, allowing them to quickly pinpoint areas requiring focused actions, thereby streamlining the decision-making process.

\par Additionally, participants valued the system's ability to provide in-depth and interactive feedback on the recommendations. The design elements, such as the sampling and fusion functions, allowed users to iteratively explore and modify the recommendation outcomes. The seamless interaction among different views enabled a smooth analytical flow, making the system highly adaptable to users' needs. This iterative examination of recommendation outcomes was particularly noted as a strength, empowering experts to dynamically adjust and refine their strategies.

\par In summary, the interface design of \textit{Prefer2SD} received high praise from users, with both visual and interaction elements achieving top ratings. The effective integration of visualization tools, along with the system's interactive and flexible design, allowed participants to conduct detailed and dynamic analyses, significantly enhancing the overall user experience.

\section{Discussion and Limitations}
\par In this section, we discuss \textit{Prefer2SD} regarding three design considerations, generalizability and scalability, and limitations and future work.

\subsection{Design Considerations}
\par \textbf{DC1: Tailored Focus on Target Users.} Previous research on recommendation systems has primarily focused on end-users, emphasizing the importance of trust-building and enhancing user experience through transparent recommendation processes~\cite{naveed2020featuristic,faltings2004designing,gretarsson2010smallworlds,bostandjiev2012tasteweights,verbert2013visualizing}. In contrast, \textit{Prefer2SD} specifically targets algorithm experts responsible for designing these recommendation systems. This strategic shift aims to enhance user experience in friend recommendations by refining the effectiveness and adaptability of recommendation algorithms. By doing so, algorithm experts can optimize user experiences for end-users by offering more diverse and flexible recommendation models. Algorithm experts should leverage through the provision of more diverse and flexible recommendation models. Algorithm experts should utilize the system's adaptability to experiment with various recommendation models and game scenarios, allowing them to refine their strategies in response to differing user behaviors and needs. Regularly reviewing user interaction data will enable them to identify patterns that inform ongoing adjustments, ensuring that the system remains responsive to evolving user requirements.

\par \textbf{DC2: Enhanced Granularity in Candidate Adjustment.} Previous approaches have typically allowed for candidate filtering across various channels~\cite{bostandjiev2012tasteweights, parra2014see}, but often lacked the fine-tuned control necessary for more nuanced recommendations. \textit{Prefer2SD} addresses this limitation by enabling experts to adjust both intra- and inter-preference ratios. This enhanced granularity provides precise control over how recommendations are customized, thereby improving flexibility and optimization potential across diverse recommendation scenarios.

For algorithm experts, it is highly recommended to utilize these granular adjustment options to fine-tune the balance between similarity and diversity, particularly in situations where subtle shifts in preferences can significantly affect recommendation quality. By closely monitoring the results of these adjustments, experts can identify the most effective combinations of ratios that resonate with different user groups.

\par \textbf{DC3: Facilitation of Iterative Group-Level Adjustments.} In contrast to prior work, \textit{Prefer2SD} integrates ratio propagation and active learning techniques, enabling experts to iteratively refine recommendations at the group level. By establishing preference ratios for a select few representative players, experts can efficiently influence recommendation outcomes for entire groups. This iterative approach not only streamlines group-level recommendations but also enhances the system's adaptability and accuracy, facilitating ongoing refinement over time. Experts are encouraged to adopt this iterative strategy, gradually adjusting group-level preferences while monitoring the collective impact on the system. Through ratio propagation techniques, they can implement changes across large groups more effectively, saving time while maintaining consistent recommendation quality across diverse user segments.

\subsection{Generalizability and Scalability}
\par This study introduces a two-step iterative and interactive framework (\cref{fig:approach overview}), aims at attaining the optimal group-level SD ratio by mediating preference ratios based on expert insight. \textbf{\textit{1) Generalizability.}} Initially, the framework is tailored for friend recommendations within MMORPGs. \xiyuan{Extending it to similar gaming contexts may require only minor adjustments, depending on the similarity of player behaviors and available data channels. For example, another MMORPG with comparable social structures would mainly involve modifying preference channels. However, in MOBA games, where player interactions are typically short-term and match-based, adapting the framework would require reconfiguring preference channels to focus on transient gameplay dynamics and immediate team composition needs.} 
Additionally, this approach represents an advancement over multi-stage recommendation systems (\cref{fig:mult-stage-rec}), theoretically adaptable to analogous recommendation processes in other domains, such as friend recommendations on social media. Nevertheless, adaptations might necessitate alterations in the channels (\cref{fig:mult-stage-rec}(A)) utilized for candidate generation, replacing or modifying the preference channels employed in this study.
\textbf{\textit{2) Scalability.}} The visualizations within the \textit{Prefer2SD} system are designed to accommodate a large volume of players. For instance, the utilization of hexbins (\cref{fig:hexbin}) in the \textit{Preference Projection View} ensures system efficiency in processing extensive player datasets. Furthermore, as our framework propagates assigned preference ratios from a few key players to the entire cohort, the efficiency and quality of \textit{Prefer2SD}'s results remain unaffected, even with larger player groups. However, \textit{Prefer2SD} may struggle with managing a multitude of preferences because visualizing each of them can lead to information overload. Thus, when utilizing \textit{Prefer2SD}, it becomes imperative to initially select a few influential preferences for analysis. \xiyuan{Additionally, the high dimensionality of combined preference vectors in multi-modal preference extraction may increase computational costs and reduce scalability, especially as the number of preferences grows.}
\label{sec:generalizability_scalability}

\subsection{Limitations and Future Work}
\par While our approach demonstrates promising capabilities, it is important to acknowledge several limitations that warrant attention.\textbf{ \textit{1) Lack of online A/B testing:}} Online A/B testing plays a pivotal role in evaluating system performance by directly measuring players' preferences for friends recommended through expert-customized strategies via \textit{Prefer2SD}. However, challenges in integrating multiple channels and navigating the complexities of the game release process have hindered the implementation of online A/B testing, resulting in its abandonment. \xiyuan{Furthermore, the lack of long-term A/B testing restricts our ability to evaluate how effectively the system maintains recommendation quality over time, including its impact on user engagement and satisfaction.} To address this gap, we conducted an quantitative within-subjects user study (\cref{sec: user_study}), partially compensating for the lack of A/B testing. By analyzing recommendation results from two perspectives - similarity to past friends and similarity to friends added in the subsequent month - we observed a significant improvement compared to the baseline, thereby partially validating the effectiveness of \textit{Prefer2SD}.
\textbf{\textit{2) Inability to handle a large number of preferences:}} As discussed in ``Scalability'' (\cref{sec:generalizability_scalability}), the current system struggles to manage an extensive array of preference settings. 
Consequently, future work should explore the development of algorithms and visualization designs capable of accommodating dozens or even hundreds of preference channels. \textbf{\textit{3) Dependence on human knowledge for recommended results:}} The current approach heavily relies on expert-selected representative players, potentially resulting in inadequate ratio propagation if the chosen players exhibit excessive similarity. 
To mitigate this issue, future iterations will integrate statistical visualization techniques to support representative selection, ensuring a more comprehensive coverage of player distribution. \textbf{\textit{4) Absence of mining for other multi-modal preferences:}} \textit{Prefer2SD} currently does not incorporate other multi-modal preference data within the game, such as text and voice data from chats on public in-game channels. While including this data may yield more accurate recommendation results, it also raises privacy concerns for players, necessitating careful consideration and addressing in future endeavors. \textbf{\textit{5) Data privacy and open-sourcing:}} Finally, while the current system remains proprietary, future work will consider open-sourcing the system after resolving data privacy concerns through discussions with the collaborating company. \xiyuan{\textbf{\textit{6) Lack of direct player involvement in the formative study.}} While expert feedback provided valuable insights into the systemic limitations, directly involving players is essential for understanding their personal preferences, which are key to the social dynamics in MMORPGs. Future work should engage a diverse range of players, including both new and experienced players, to ensure the design requirements are closely aligned with their needs. \textbf{\textit{7) Cold-start problem.}} The current design does not explicitly tackle the cold-start problem for new players with limited or no historical data. Further iterations of \textit{Prefer2SD} could address this issue by incorporating methods such as demographic-based initialization or collaborative strategies for small datasets. \textit{\textbf{8) Computational expense of ratio propagation.}} Ratio propagation can become computationally expensive when scaling to tens of thousands of players due to increased computational load. To address this, we plan to implement parallel computing techniques to enhance computational efficiency. \textbf{\textit{9) Limited sample size in the user study.}} The user study included only 12 participants, which may limit the generalizability of the findings. Future research should involve a larger and more diverse group of experts to ensure the system's design is applicable across different domains.}

\section{Conclusion}
\par This study introduces \textit{Prefer2SD}, a two-step human-in-the-loop workflow empowered by an interactive visual analytics system. This system aids algorithm experts in navigating, evaluating, and harmonizing the similarity and diversity in recommended friends for specific player groups. Our method integrates a range of player preferences and provides carefully designed interactive visualizations to assist experts in iteratively adjusting preference ratios, striving for an ideal equilibrium between similarity and diversity. The effectiveness of \textit{Prefer2SD} is substantiated through a within-subjects study, a case study, and expert interviews, affirming its practicality and relevance.

\begin{acks}
\par We thank anonymous reviewers for their valuable feedback. This work is supported by grants from the National Natural Science Foundation of China (No. 62372298), Shanghai Engineering Research Center of Intelligent Vision and Imaging, Shanghai Frontiers Science Center of Human-centered Artificial Intelligence (ShangHAI), and MoE Key Laboratory of Intelligent Perception and Human-Machine Collaboration (KLIP-HuMaCo).
\end{acks}

\bibliographystyle{ACM-Reference-Format}
\bibliography{sample-base}


\begin{thebibliography}{102}


\ifx \showCODEN    \undefined \def \showCODEN     #1{\unskip}     \fi
\ifx \showDOI      \undefined \def \showDOI       #1{#1}\fi
\ifx \showISBNx    \undefined \def \showISBNx     #1{\unskip}     \fi
\ifx \showISBNxiii \undefined \def \showISBNxiii  #1{\unskip}     \fi
\ifx \showISSN     \undefined \def \showISSN      #1{\unskip}     \fi
\ifx \showLCCN     \undefined \def \showLCCN      #1{\unskip}     \fi
\ifx \shownote     \undefined \def \shownote      #1{#1}          \fi
\ifx \showarticletitle \undefined \def \showarticletitle #1{#1}   \fi
\ifx \showURL      \undefined \def \showURL       {\relax}        \fi
\providecommand\bibfield[2]{#2}
\providecommand\bibinfo[2]{#2}
\providecommand\natexlab[1]{#1}
\providecommand\showeprint[2][]{arXiv:#2}

\bibitem[\protect\citeauthoryear{Adomavicius and Kwon}{Adomavicius and
  Kwon}{2011}]%
        {adomavicius2011improving}
\bibfield{author}{\bibinfo{person}{Gediminas Adomavicius} {and}
  \bibinfo{person}{YoungOk Kwon}.} \bibinfo{year}{2011}\natexlab{}.
\newblock \showarticletitle{Improving aggregate recommendation diversity using
  ranking-based techniques}.
\newblock \bibinfo{journal}{\emph{IEEE Transactions on Knowledge and Data
  Engineering}} \bibinfo{volume}{24}, \bibinfo{number}{5}
  (\bibinfo{year}{2011}), \bibinfo{pages}{896--911}.
\newblock


\bibitem[\protect\citeauthoryear{Agarwal and Bharadwaj}{Agarwal and
  Bharadwaj}{2013}]%
        {agarwal2013collaborative}
\bibfield{author}{\bibinfo{person}{Vinti Agarwal} {and}
  \bibinfo{person}{Kamal~Kant Bharadwaj}.} \bibinfo{year}{2013}\natexlab{}.
\newblock \showarticletitle{A collaborative filtering framework for friends
  recommendation in social networks based on interaction intensity and adaptive
  user similarity}.
\newblock \bibinfo{journal}{\emph{Social Network Analysis and Mining}}
  \bibinfo{volume}{3} (\bibinfo{year}{2013}), \bibinfo{pages}{359--379}.
\newblock


\bibitem[\protect\citeauthoryear{Ak, Schifferer, Rabhi, and De~Souza
  Pereira~Moreira}{Ak et~al\mbox{.}}{2022}]%
        {ak2022training}
\bibfield{author}{\bibinfo{person}{Ronay Ak}, \bibinfo{person}{Benedikt
  Schifferer}, \bibinfo{person}{Sara Rabhi}, {and} \bibinfo{person}{Gabriel
  De~Souza Pereira~Moreira}.} \bibinfo{year}{2022}\natexlab{}.
\newblock \showarticletitle{Training and Deploying Multi-Stage Recommender
  Systems}. In \bibinfo{booktitle}{\emph{Proceedings of the 16th ACM Conference
  on Recommender Systems}}. \bibinfo{pages}{706--707}.
\newblock


\bibitem[\protect\citeauthoryear{Algarni and Sheldon}{Algarni and
  Sheldon}{2023}]%
        {algarni2023systematic}
\bibfield{author}{\bibinfo{person}{Shrooq Algarni} {and}
  \bibinfo{person}{Frederick Sheldon}.} \bibinfo{year}{2023}\natexlab{}.
\newblock \showarticletitle{Systematic Review of Recommendation Systems for
  Course Selection}.
\newblock \bibinfo{journal}{\emph{Machine Learning and Knowledge Extraction}}
  \bibinfo{volume}{5}, \bibinfo{number}{2} (\bibinfo{year}{2023}),
  \bibinfo{pages}{560--596}.
\newblock


\bibitem[\protect\citeauthoryear{Andjelkovic, Parra, and O'Donovan}{Andjelkovic
  et~al\mbox{.}}{2016}]%
        {andjelkovic2016moodplay}
\bibfield{author}{\bibinfo{person}{Ivana Andjelkovic}, \bibinfo{person}{Denis
  Parra}, {and} \bibinfo{person}{John O'Donovan}.}
  \bibinfo{year}{2016}\natexlab{}.
\newblock \showarticletitle{Moodplay: Interactive mood-based music discovery
  and recommendation}. In \bibinfo{booktitle}{\emph{Proceedings of the 2016
  conference on user modeling adaptation and personalization}}.
  \bibinfo{pages}{275--279}.
\newblock
\urldef\tempurl%
\url{https://doi.org/10.1145/2930238.2930280}
\showDOI{\tempurl}


\bibitem[\protect\citeauthoryear{Andjelkovic, Parra, and
  O’Donovan}{Andjelkovic et~al\mbox{.}}{2019}]%
        {andjelkovic2019moodplay}
\bibfield{author}{\bibinfo{person}{Ivana Andjelkovic}, \bibinfo{person}{Denis
  Parra}, {and} \bibinfo{person}{John O’Donovan}.}
  \bibinfo{year}{2019}\natexlab{}.
\newblock \showarticletitle{Moodplay: interactive music recommendation based on
  artists’ mood similarity}.
\newblock \bibinfo{journal}{\emph{International Journal of Human-Computer
  Studies}}  \bibinfo{volume}{121} (\bibinfo{year}{2019}),
  \bibinfo{pages}{142--159}.
\newblock


\bibitem[\protect\citeauthoryear{Ashktorab, Desmond, Andres, Muller, Joshi,
  Brachman, Sharma, Brimijoin, Pan, Wolf, et~al\mbox{.}}{Ashktorab
  et~al\mbox{.}}{2021}]%
        {ashktorab2021ai}
\bibfield{author}{\bibinfo{person}{Zahra Ashktorab}, \bibinfo{person}{Michael
  Desmond}, \bibinfo{person}{Josh Andres}, \bibinfo{person}{Michael Muller},
  \bibinfo{person}{Narendra~Nath Joshi}, \bibinfo{person}{Michelle Brachman},
  \bibinfo{person}{Aabhas Sharma}, \bibinfo{person}{Kristina Brimijoin},
  \bibinfo{person}{Qian Pan}, \bibinfo{person}{Christine~T Wolf},
  {et~al\mbox{.}}} \bibinfo{year}{2021}\natexlab{}.
\newblock \showarticletitle{Ai-assisted human labeling: Batching for efficiency
  without overreliance}.
\newblock \bibinfo{journal}{\emph{Proceedings of the ACM on Human-Computer
  Interaction}} \bibinfo{volume}{5}, \bibinfo{number}{CSCW1}
  (\bibinfo{year}{2021}), \bibinfo{pages}{1--27}.
\newblock


\bibitem[\protect\citeauthoryear{Bagci and Karagoz}{Bagci and Karagoz}{2016}]%
        {bagci2016context}
\bibfield{author}{\bibinfo{person}{Hakan Bagci} {and} \bibinfo{person}{Pinar
  Karagoz}.} \bibinfo{year}{2016}\natexlab{}.
\newblock \showarticletitle{Context-aware friend recommendation for location
  based social networks using random walk}. In
  \bibinfo{booktitle}{\emph{Proceedings of the 25th international conference
  companion on world wide web}}. \bibinfo{pages}{531--536}.
\newblock


\bibitem[\protect\citeauthoryear{Bangor, Kortum, and Miller}{Bangor
  et~al\mbox{.}}{2009}]%
        {bangor2009determining}
\bibfield{author}{\bibinfo{person}{Aaron Bangor}, \bibinfo{person}{Philip
  Kortum}, {and} \bibinfo{person}{James Miller}.}
  \bibinfo{year}{2009}\natexlab{}.
\newblock \showarticletitle{Determining what individual SUS scores mean: Adding
  an adjective rating scale}.
\newblock \bibinfo{journal}{\emph{Journal of usability studies}}
  \bibinfo{volume}{4}, \bibinfo{number}{3} (\bibinfo{year}{2009}),
  \bibinfo{pages}{114--123}.
\newblock


\bibitem[\protect\citeauthoryear{Berkhin}{Berkhin}{2005}]%
        {berkhin2005survey}
\bibfield{author}{\bibinfo{person}{Pavel Berkhin}.}
  \bibinfo{year}{2005}\natexlab{}.
\newblock \showarticletitle{A survey on PageRank computing}.
\newblock \bibinfo{journal}{\emph{Internet mathematics}} \bibinfo{volume}{2},
  \bibinfo{number}{1} (\bibinfo{year}{2005}), \bibinfo{pages}{73--120}.
\newblock


\bibitem[\protect\citeauthoryear{Bernard, Zeppelzauer, Sedlmair, and
  Aigner}{Bernard et~al\mbox{.}}{2018}]%
        {bernard2018vial}
\bibfield{author}{\bibinfo{person}{J{\"u}rgen Bernard},
  \bibinfo{person}{Matthias Zeppelzauer}, \bibinfo{person}{Michael Sedlmair},
  {and} \bibinfo{person}{Wolfgang Aigner}.} \bibinfo{year}{2018}\natexlab{}.
\newblock \showarticletitle{VIAL: a unified process for visual interactive
  labeling}.
\newblock \bibinfo{journal}{\emph{The Visual Computer}} \bibinfo{volume}{34},
  \bibinfo{number}{9} (\bibinfo{year}{2018}), \bibinfo{pages}{1189--1207}.
\newblock


\bibitem[\protect\citeauthoryear{Bian and Holtzman}{Bian and Holtzman}{2011}]%
        {bian2011online}
\bibfield{author}{\bibinfo{person}{Li Bian} {and} \bibinfo{person}{Henry
  Holtzman}.} \bibinfo{year}{2011}\natexlab{}.
\newblock \showarticletitle{Online friend recommendation through personality
  matching and collaborative filtering}.
\newblock \bibinfo{journal}{\emph{Proc. of UBICOMM}} \bibinfo{volume}{5},
  \bibinfo{number}{2011} (\bibinfo{year}{2011}), \bibinfo{pages}{230--235}.
\newblock


\bibitem[\protect\citeauthoryear{Blandford, Furniss, and Makri}{Blandford
  et~al\mbox{.}}{2016}]%
        {blandford2016qualitative}
\bibfield{author}{\bibinfo{person}{Ann Blandford}, \bibinfo{person}{Dominic
  Furniss}, {and} \bibinfo{person}{Stephann Makri}.}
  \bibinfo{year}{2016}\natexlab{}.
\newblock \bibinfo{booktitle}{\emph{Qualitative HCI research: Going behind the
  scenes}}.
\newblock \bibinfo{publisher}{Morgan \& Claypool Publishers}.
\newblock


\bibitem[\protect\citeauthoryear{Bostandjiev, O'Donovan, and
  H{\"o}llerer}{Bostandjiev et~al\mbox{.}}{2012}]%
        {bostandjiev2012tasteweights}
\bibfield{author}{\bibinfo{person}{Svetlin Bostandjiev}, \bibinfo{person}{John
  O'Donovan}, {and} \bibinfo{person}{Tobias H{\"o}llerer}.}
  \bibinfo{year}{2012}\natexlab{}.
\newblock \showarticletitle{TasteWeights: a visual interactive hybrid
  recommender system}. In \bibinfo{booktitle}{\emph{Proceedings of the sixth
  ACM conference on Recommender systems}}. \bibinfo{pages}{35--42}.
\newblock


\bibitem[\protect\citeauthoryear{Brachman, Ashktorab, Desmond, Duesterwald,
  Dugan, Joshi, Pan, and Sharma}{Brachman et~al\mbox{.}}{2022}]%
        {brachman2022reliance}
\bibfield{author}{\bibinfo{person}{Michelle Brachman}, \bibinfo{person}{Zahra
  Ashktorab}, \bibinfo{person}{Michael Desmond}, \bibinfo{person}{Evelyn
  Duesterwald}, \bibinfo{person}{Casey Dugan}, \bibinfo{person}{Narendra~Nath
  Joshi}, \bibinfo{person}{Qian Pan}, {and} \bibinfo{person}{Aabhas Sharma}.}
  \bibinfo{year}{2022}\natexlab{}.
\newblock \showarticletitle{Reliance and Automation for Human-AI Collaborative
  Data Labeling Conflict Resolution}.
\newblock \bibinfo{journal}{\emph{Proceedings of the ACM on Human-Computer
  Interaction}} \bibinfo{volume}{6}, \bibinfo{number}{CSCW2}
  (\bibinfo{year}{2022}), \bibinfo{pages}{1--27}.
\newblock


\bibitem[\protect\citeauthoryear{Brame}{Brame}{2016}]%
        {brame2016active}
\bibfield{author}{\bibinfo{person}{Cynthia Brame}.}
  \bibinfo{year}{2016}\natexlab{}.
\newblock \showarticletitle{Active learning}.
\newblock \bibinfo{journal}{\emph{Vanderbilt University Center for Teaching}}
  (\bibinfo{year}{2016}).
\newblock


\bibitem[\protect\citeauthoryear{Braun and Clarke}{Braun and Clarke}{2012}]%
        {braun2012thematic}
\bibfield{author}{\bibinfo{person}{Virginia Braun} {and}
  \bibinfo{person}{Victoria Clarke}.} \bibinfo{year}{2012}\natexlab{}.
\newblock \bibinfo{booktitle}{\emph{Thematic analysis.}}
\newblock \bibinfo{publisher}{American Psychological Association}.
\newblock


\bibitem[\protect\citeauthoryear{Carpenter}{Carpenter}{2010}]%
        {carpenter2010study}
\bibfield{author}{\bibinfo{person}{Serena Carpenter}.}
  \bibinfo{year}{2010}\natexlab{}.
\newblock \showarticletitle{A study of content diversity in online citizen
  journalism and online newspaper articles}.
\newblock \bibinfo{journal}{\emph{New media \& society}} \bibinfo{volume}{12},
  \bibinfo{number}{7} (\bibinfo{year}{2010}), \bibinfo{pages}{1064--1084}.
\newblock


\bibitem[\protect\citeauthoryear{Chang and Lin}{Chang and Lin}{2014}]%
        {chang2014team}
\bibfield{author}{\bibinfo{person}{Shan-Mei Chang} {and}
  \bibinfo{person}{Sunny~SJ Lin}.} \bibinfo{year}{2014}\natexlab{}.
\newblock \showarticletitle{Team knowledge with motivation in a successful
  MMORPG game team: A case study}.
\newblock \bibinfo{journal}{\emph{Computers \& Education}}
  \bibinfo{volume}{73} (\bibinfo{year}{2014}), \bibinfo{pages}{129--140}.
\newblock


\bibitem[\protect\citeauthoryear{Chang, Shu, Du, Chen, Zhang, Zheng, Huang, and
  Xing}{Chang et~al\mbox{.}}{2022}]%
        {chang2022graphrr}
\bibfield{author}{\bibinfo{person}{Yaomin Chang}, \bibinfo{person}{Lin Shu},
  \bibinfo{person}{Erxin Du}, \bibinfo{person}{Chuan Chen},
  \bibinfo{person}{Ziyang Zhang}, \bibinfo{person}{Zibin Zheng},
  \bibinfo{person}{Yuzhao Huang}, {and} \bibinfo{person}{Xingxing Xing}.}
  \bibinfo{year}{2022}\natexlab{}.
\newblock \showarticletitle{GraphRR: A multiplex Graph based Reciprocal friend
  Recommender system with applications on online gaming service}.
\newblock \bibinfo{journal}{\emph{Knowledge-Based Systems}}
  \bibinfo{volume}{251} (\bibinfo{year}{2022}), \bibinfo{pages}{109187}.
\newblock
\urldef\tempurl%
\url{https://doi.org/10.1016/j.knosys.2022.109187}
\showDOI{\tempurl}


\bibitem[\protect\citeauthoryear{Chen, Xie, Zheng, Zheng, and Xie}{Chen
  et~al\mbox{.}}{2020}]%
        {chen2020friend}
\bibfield{author}{\bibinfo{person}{Liang Chen}, \bibinfo{person}{Yuanzhen Xie},
  \bibinfo{person}{Zibin Zheng}, \bibinfo{person}{Huayou Zheng}, {and}
  \bibinfo{person}{Jingdun Xie}.} \bibinfo{year}{2020}\natexlab{}.
\newblock \showarticletitle{Friend recommendation based on multi-social graph
  convolutional network}.
\newblock \bibinfo{journal}{\emph{IEEE Access}}  \bibinfo{volume}{8}
  (\bibinfo{year}{2020}), \bibinfo{pages}{43618--43629}.
\newblock


\bibitem[\protect\citeauthoryear{Chen, Zeng, and Yuan}{Chen
  et~al\mbox{.}}{2013}]%
        {chen2013unified}
\bibfield{author}{\bibinfo{person}{Li Chen}, \bibinfo{person}{Wei Zeng}, {and}
  \bibinfo{person}{Quan Yuan}.} \bibinfo{year}{2013}\natexlab{}.
\newblock \showarticletitle{A unified framework for recommending items, groups
  and friends in social media environment via mutual resource fusion}.
\newblock \bibinfo{journal}{\emph{Expert Systems with Applications}}
  \bibinfo{volume}{40}, \bibinfo{number}{8} (\bibinfo{year}{2013}),
  \bibinfo{pages}{2889--2903}.
\newblock


\bibitem[\protect\citeauthoryear{Cheng, Zhang, Zou, Huang, and Zhang}{Cheng
  et~al\mbox{.}}{2019}]%
        {cheng2019friend}
\bibfield{author}{\bibinfo{person}{Shulin Cheng}, \bibinfo{person}{Bofeng
  Zhang}, \bibinfo{person}{Guobing Zou}, \bibinfo{person}{Mingqing Huang},
  {and} \bibinfo{person}{Zhu Zhang}.} \bibinfo{year}{2019}\natexlab{}.
\newblock \showarticletitle{Friend recommendation in social networks based on
  multi-source information fusion}.
\newblock \bibinfo{journal}{\emph{International Journal of Machine Learning and
  Cybernetics}}  \bibinfo{volume}{10} (\bibinfo{year}{2019}),
  \bibinfo{pages}{1003--1024}.
\newblock


\bibitem[\protect\citeauthoryear{Choi, Park, Yang, Kim, Choo, and Hong}{Choi
  et~al\mbox{.}}{2019}]%
        {choi2019aila}
\bibfield{author}{\bibinfo{person}{Minsuk Choi}, \bibinfo{person}{Cheonbok
  Park}, \bibinfo{person}{Soyoung Yang}, \bibinfo{person}{Yonggyu Kim},
  \bibinfo{person}{Jaegul Choo}, {and} \bibinfo{person}{Sungsoo~Ray Hong}.}
  \bibinfo{year}{2019}\natexlab{}.
\newblock \showarticletitle{Aila: Attentive interactive labeling assistant for
  document classification through attention-based deep neural networks}. In
  \bibinfo{booktitle}{\emph{Proceedings of the 2019 CHI conference on human
  factors in computing systems}}. \bibinfo{pages}{1--12}.
\newblock


\bibitem[\protect\citeauthoryear{Claypool and Claypool}{Claypool and
  Claypool}{2007}]%
        {claypool2007frame}
\bibfield{author}{\bibinfo{person}{Kajal~T Claypool} {and}
  \bibinfo{person}{Mark Claypool}.} \bibinfo{year}{2007}\natexlab{}.
\newblock \showarticletitle{On frame rate and player performance in first
  person shooter games}.
\newblock \bibinfo{journal}{\emph{Multimedia systems}} \bibinfo{volume}{13},
  \bibinfo{number}{1} (\bibinfo{year}{2007}), \bibinfo{pages}{3--17}.
\newblock


\bibitem[\protect\citeauthoryear{Dorogovtsev, Goltsev, and Mendes}{Dorogovtsev
  et~al\mbox{.}}{2006}]%
        {dorogovtsev2006k}
\bibfield{author}{\bibinfo{person}{Sergey~N Dorogovtsev},
  \bibinfo{person}{Alexander~V Goltsev}, {and} \bibinfo{person}{Jose Ferreira~F
  Mendes}.} \bibinfo{year}{2006}\natexlab{}.
\newblock \showarticletitle{K-core organization of complex networks}.
\newblock \bibinfo{journal}{\emph{Physical review letters}}
  \bibinfo{volume}{96}, \bibinfo{number}{4} (\bibinfo{year}{2006}),
  \bibinfo{pages}{040601}.
\newblock
\urldef\tempurl%
\url{https://doi.org/10.1007/978-1-4614-0754-6_9}
\showDOI{\tempurl}


\bibitem[\protect\citeauthoryear{Druck, Settles, and McCallum}{Druck
  et~al\mbox{.}}{2009}]%
        {druck2009active}
\bibfield{author}{\bibinfo{person}{Gregory Druck}, \bibinfo{person}{Burr
  Settles}, {and} \bibinfo{person}{Andrew McCallum}.}
  \bibinfo{year}{2009}\natexlab{}.
\newblock \showarticletitle{Active learning by labeling features}. In
  \bibinfo{booktitle}{\emph{Proceedings of the 2009 conference on Empirical
  methods in natural language processing}}. \bibinfo{pages}{81--90}.
\newblock
\urldef\tempurl%
\url{https://doi.org/10.3115/1699510.1699522}
\showDOI{\tempurl}


\bibitem[\protect\citeauthoryear{Duell}{Duell}{2014}]%
        {duell2014team}
\bibfield{author}{\bibinfo{person}{Adam Duell}.}
  \bibinfo{year}{2014}\natexlab{}.
\newblock \showarticletitle{From team play to squad play: The militarisation of
  interactions in multiplayer FPS video games}.
\newblock \bibinfo{journal}{\emph{Press Start}} \bibinfo{volume}{1},
  \bibinfo{number}{1} (\bibinfo{year}{2014}), \bibinfo{pages}{59--78}.
\newblock


\bibitem[\protect\citeauthoryear{Eskandanian and Mobasher}{Eskandanian and
  Mobasher}{2020}]%
        {eskandanian2020using}
\bibfield{author}{\bibinfo{person}{Farzad Eskandanian} {and}
  \bibinfo{person}{Bamshad Mobasher}.} \bibinfo{year}{2020}\natexlab{}.
\newblock \showarticletitle{Using stable matching to optimize the balance
  between accuracy and diversity in recommendation}. In
  \bibinfo{booktitle}{\emph{Proceedings of the 28th ACM Conference on User
  Modeling, Adaptation and Personalization}}. \bibinfo{pages}{71--79}.
\newblock


\bibitem[\protect\citeauthoryear{Faltings, Pu, Torrens, and Viappiani}{Faltings
  et~al\mbox{.}}{2004}]%
        {faltings2004designing}
\bibfield{author}{\bibinfo{person}{Boi Faltings}, \bibinfo{person}{Pearl Pu},
  \bibinfo{person}{Marc Torrens}, {and} \bibinfo{person}{Paolo Viappiani}.}
  \bibinfo{year}{2004}\natexlab{}.
\newblock \showarticletitle{Designing example-critiquing interaction}. In
  \bibinfo{booktitle}{\emph{Proceedings of the 9th international conference on
  Intelligent user interfaces}}. \bibinfo{pages}{22--29}.
\newblock


\bibitem[\protect\citeauthoryear{Fan, Ma, Li, He, Zhao, Tang, and Yin}{Fan
  et~al\mbox{.}}{2019}]%
        {fan2019graph}
\bibfield{author}{\bibinfo{person}{Wenqi Fan}, \bibinfo{person}{Yao Ma},
  \bibinfo{person}{Qing Li}, \bibinfo{person}{Yuan He}, \bibinfo{person}{Eric
  Zhao}, \bibinfo{person}{Jiliang Tang}, {and} \bibinfo{person}{Dawei Yin}.}
  \bibinfo{year}{2019}\natexlab{}.
\newblock \showarticletitle{Graph neural networks for social recommendation}.
  In \bibinfo{booktitle}{\emph{The world wide web conference}}.
  \bibinfo{pages}{417--426}.
\newblock


\bibitem[\protect\citeauthoryear{Feng, Yu, and Zhou}{Feng
  et~al\mbox{.}}{2018}]%
        {feng2018multi}
\bibfield{author}{\bibinfo{person}{Ji Feng}, \bibinfo{person}{Yang Yu}, {and}
  \bibinfo{person}{Zhi-Hua Zhou}.} \bibinfo{year}{2018}\natexlab{}.
\newblock \showarticletitle{Multi-layered gradient boosting decision trees}.
\newblock \bibinfo{journal}{\emph{Advances in neural information processing
  systems}}  \bibinfo{volume}{31} (\bibinfo{year}{2018}).
\newblock


\bibitem[\protect\citeauthoryear{Fkih}{Fkih}{2022}]%
        {fkih2022similarity}
\bibfield{author}{\bibinfo{person}{Fethi Fkih}.}
  \bibinfo{year}{2022}\natexlab{}.
\newblock \showarticletitle{Similarity measures for Collaborative
  Filtering-based Recommender Systems: Review and experimental comparison}.
\newblock \bibinfo{journal}{\emph{Journal of King Saud University-Computer and
  Information Sciences}} \bibinfo{volume}{34}, \bibinfo{number}{9}
  (\bibinfo{year}{2022}), \bibinfo{pages}{7645--7669}.
\newblock


\bibitem[\protect\citeauthoryear{Gao, Zheng, Li, Li, Qin, Piao, Quan, Chang,
  Jin, He, et~al\mbox{.}}{Gao et~al\mbox{.}}{2023}]%
        {gao2023survey}
\bibfield{author}{\bibinfo{person}{Chen Gao}, \bibinfo{person}{Yu Zheng},
  \bibinfo{person}{Nian Li}, \bibinfo{person}{Yinfeng Li},
  \bibinfo{person}{Yingrong Qin}, \bibinfo{person}{Jinghua Piao},
  \bibinfo{person}{Yuhan Quan}, \bibinfo{person}{Jianxin Chang},
  \bibinfo{person}{Depeng Jin}, \bibinfo{person}{Xiangnan He}, {et~al\mbox{.}}}
  \bibinfo{year}{2023}\natexlab{}.
\newblock \showarticletitle{A survey of graph neural networks for recommender
  systems: Challenges, methods, and directions}.
\newblock \bibinfo{journal}{\emph{ACM Transactions on Recommender Systems}}
  \bibinfo{volume}{1}, \bibinfo{number}{1} (\bibinfo{year}{2023}),
  \bibinfo{pages}{1--51}.
\newblock


\bibitem[\protect\citeauthoryear{Glickman}{Glickman}{1995}]%
        {glickman1995glicko}
\bibfield{author}{\bibinfo{person}{Mark~E Glickman}.}
  \bibinfo{year}{1995}\natexlab{}.
\newblock \showarticletitle{The glicko system}.
\newblock \bibinfo{journal}{\emph{Boston University}} \bibinfo{volume}{16},
  \bibinfo{number}{8} (\bibinfo{year}{1995}), \bibinfo{pages}{9}.
\newblock


\bibitem[\protect\citeauthoryear{Gratzl, Lex, Gehlenborg, Pfister, and
  Streit}{Gratzl et~al\mbox{.}}{2013}]%
        {gratzl2013lineup}
\bibfield{author}{\bibinfo{person}{Samuel Gratzl}, \bibinfo{person}{Alexander
  Lex}, \bibinfo{person}{Nils Gehlenborg}, \bibinfo{person}{Hanspeter Pfister},
  {and} \bibinfo{person}{Marc Streit}.} \bibinfo{year}{2013}\natexlab{}.
\newblock \showarticletitle{Lineup: Visual analysis of multi-attribute
  rankings}.
\newblock \bibinfo{journal}{\emph{IEEE transactions on visualization and
  computer graphics}} \bibinfo{volume}{19}, \bibinfo{number}{12}
  (\bibinfo{year}{2013}), \bibinfo{pages}{2277--2286}.
\newblock
\urldef\tempurl%
\url{https://doi.org/10.1109/tvcg.2013.173}
\showDOI{\tempurl}


\bibitem[\protect\citeauthoryear{Gretarsson, O'Donovan, Bostandjiev, Hall, and
  H{\"o}llerer}{Gretarsson et~al\mbox{.}}{2010}]%
        {gretarsson2010smallworlds}
\bibfield{author}{\bibinfo{person}{Brynjar Gretarsson}, \bibinfo{person}{John
  O'Donovan}, \bibinfo{person}{Svetlin Bostandjiev},
  \bibinfo{person}{Christopher Hall}, {and} \bibinfo{person}{Tobias
  H{\"o}llerer}.} \bibinfo{year}{2010}\natexlab{}.
\newblock \showarticletitle{Smallworlds: visualizing social recommendations}.
  In \bibinfo{booktitle}{\emph{Computer graphics forum}},
  Vol.~\bibinfo{volume}{29}. Wiley Online Library, \bibinfo{pages}{833--842}.
\newblock


\bibitem[\protect\citeauthoryear{Grover and Leskovec}{Grover and
  Leskovec}{2016}]%
        {grover2016node2vec}
\bibfield{author}{\bibinfo{person}{Aditya Grover} {and} \bibinfo{person}{Jure
  Leskovec}.} \bibinfo{year}{2016}\natexlab{}.
\newblock \showarticletitle{node2vec: Scalable feature learning for networks}.
  In \bibinfo{booktitle}{\emph{Proceedings of the 22nd ACM SIGKDD international
  conference on Knowledge discovery and data mining}}.
  \bibinfo{pages}{855--864}.
\newblock


\bibitem[\protect\citeauthoryear{Hart}{Hart}{2006}]%
        {hart2006nasa}
\bibfield{author}{\bibinfo{person}{Sandra~G Hart}.}
  \bibinfo{year}{2006}\natexlab{}.
\newblock \showarticletitle{NASA-task load index (NASA-TLX); 20 years later}.
  In \bibinfo{booktitle}{\emph{Proceedings of the human factors and ergonomics
  society annual meeting}}, Vol.~\bibinfo{volume}{50}. Sage publications Sage
  CA: Los Angeles, CA, \bibinfo{pages}{904--908}.
\newblock


\bibitem[\protect\citeauthoryear{Hinds, Carley, Krackhardt, and Wholey}{Hinds
  et~al\mbox{.}}{2000}]%
        {hinds2000choosing}
\bibfield{author}{\bibinfo{person}{Pamela~J Hinds}, \bibinfo{person}{Kathleen~M
  Carley}, \bibinfo{person}{David Krackhardt}, {and} \bibinfo{person}{Doug
  Wholey}.} \bibinfo{year}{2000}\natexlab{}.
\newblock \showarticletitle{Choosing work group members: Balancing similarity,
  competence, and familiarity}.
\newblock \bibinfo{journal}{\emph{Organizational behavior and human decision
  processes}} \bibinfo{volume}{81}, \bibinfo{number}{2} (\bibinfo{year}{2000}),
  \bibinfo{pages}{226--251}.
\newblock


\bibitem[\protect\citeauthoryear{Huang, Zhang, Wang, and Hua}{Huang
  et~al\mbox{.}}{2015}]%
        {huang2015social}
\bibfield{author}{\bibinfo{person}{Shangrong Huang}, \bibinfo{person}{Jian
  Zhang}, \bibinfo{person}{Lei Wang}, {and} \bibinfo{person}{Xian-Sheng Hua}.}
  \bibinfo{year}{2015}\natexlab{}.
\newblock \showarticletitle{Social friend recommendation based on multiple
  network correlation}.
\newblock \bibinfo{journal}{\emph{IEEE transactions on multimedia}}
  \bibinfo{volume}{18}, \bibinfo{number}{2} (\bibinfo{year}{2015}),
  \bibinfo{pages}{287--299}.
\newblock


\bibitem[\protect\citeauthoryear{Iscen, Tolias, Avrithis, and Chum}{Iscen
  et~al\mbox{.}}{2019}]%
        {iscen2019label}
\bibfield{author}{\bibinfo{person}{Ahmet Iscen}, \bibinfo{person}{Giorgos
  Tolias}, \bibinfo{person}{Yannis Avrithis}, {and} \bibinfo{person}{Ondrej
  Chum}.} \bibinfo{year}{2019}\natexlab{}.
\newblock \showarticletitle{Label propagation for deep semi-supervised
  learning}. In \bibinfo{booktitle}{\emph{Proceedings of the IEEE/CVF
  conference on computer vision and pattern recognition}}.
  \bibinfo{pages}{5070--5079}.
\newblock


\bibitem[\protect\citeauthoryear{J{\"a}{\"a}skel{\"a}inen}{J{\"a}{\"a}skel{\"a}inen}{2010}]%
        {jaaskelainen2010think}
\bibfield{author}{\bibinfo{person}{Riitta J{\"a}{\"a}skel{\"a}inen}.}
  \bibinfo{year}{2010}\natexlab{}.
\newblock \showarticletitle{Think-aloud protocol}.
\newblock \bibinfo{journal}{\emph{Handbook of translation studies}}
  \bibinfo{volume}{1} (\bibinfo{year}{2010}), \bibinfo{pages}{371--374}.
\newblock


\bibitem[\protect\citeauthoryear{Jin, Seipp, Duval, and Verbert}{Jin
  et~al\mbox{.}}{2016}]%
        {jin2016go}
\bibfield{author}{\bibinfo{person}{Yucheng Jin}, \bibinfo{person}{Karsten
  Seipp}, \bibinfo{person}{Erik Duval}, {and} \bibinfo{person}{Katrien
  Verbert}.} \bibinfo{year}{2016}\natexlab{}.
\newblock \showarticletitle{Go with the flow: effects of transparency and user
  control on targeted advertising using flow charts}. In
  \bibinfo{booktitle}{\emph{Proceedings of the international working conference
  on advanced visual interfaces}}. \bibinfo{pages}{68--75}.
\newblock


\bibitem[\protect\citeauthoryear{Ke, Meng, Finley, Wang, Chen, Ma, Ye, and
  Liu}{Ke et~al\mbox{.}}{2017}]%
        {ke2017lightgbm}
\bibfield{author}{\bibinfo{person}{Guolin Ke}, \bibinfo{person}{Qi Meng},
  \bibinfo{person}{Thomas Finley}, \bibinfo{person}{Taifeng Wang},
  \bibinfo{person}{Wei Chen}, \bibinfo{person}{Weidong Ma},
  \bibinfo{person}{Qiwei Ye}, {and} \bibinfo{person}{Tie-Yan Liu}.}
  \bibinfo{year}{2017}\natexlab{}.
\newblock \showarticletitle{Lightgbm: A highly efficient gradient boosting
  decision tree}.
\newblock \bibinfo{journal}{\emph{Advances in neural information processing
  systems}}  \bibinfo{volume}{30} (\bibinfo{year}{2017}).
\newblock


\bibitem[\protect\citeauthoryear{Kim, Li, and Hemphill}{Kim
  et~al\mbox{.}}{2024}]%
        {kim2024communication}
\bibfield{author}{\bibinfo{person}{Ji~Eun Kim}, \bibinfo{person}{Lingyao Li},
  {and} \bibinfo{person}{Libby Hemphill}.} \bibinfo{year}{2024}\natexlab{}.
\newblock \showarticletitle{Communication strategies for improving performance
  in virtual teams: Lessons from Dota 2}.
\newblock \bibinfo{journal}{\emph{Authorea Preprints}} (\bibinfo{year}{2024}).
\newblock


\bibitem[\protect\citeauthoryear{Klimashevskaia, Elahi, Jannach, Trattner, and
  Skj{\ae}rven}{Klimashevskaia et~al\mbox{.}}{2022}]%
        {klimashevskaia2022mitigating}
\bibfield{author}{\bibinfo{person}{Anastasiia Klimashevskaia},
  \bibinfo{person}{Mehdi Elahi}, \bibinfo{person}{Dietmar Jannach},
  \bibinfo{person}{Christoph Trattner}, {and} \bibinfo{person}{Lars
  Skj{\ae}rven}.} \bibinfo{year}{2022}\natexlab{}.
\newblock \showarticletitle{Mitigating popularity bias in recommendation:
  Potential and limits of calibration approaches}. In
  \bibinfo{booktitle}{\emph{International Workshop on Algorithmic Bias in
  Search and Recommendation}}. Springer, \bibinfo{pages}{82--90}.
\newblock


\bibitem[\protect\citeauthoryear{Kunaver and Po{\v{z}}rl}{Kunaver and
  Po{\v{z}}rl}{2017}]%
        {kunaver2017diversity}
\bibfield{author}{\bibinfo{person}{Matev{\v{z}} Kunaver} {and}
  \bibinfo{person}{Toma{\v{z}} Po{\v{z}}rl}.} \bibinfo{year}{2017}\natexlab{}.
\newblock \showarticletitle{Diversity in recommender systems--A survey}.
\newblock \bibinfo{journal}{\emph{Knowledge-based systems}}
  \bibinfo{volume}{123} (\bibinfo{year}{2017}), \bibinfo{pages}{154--162}.
\newblock


\bibitem[\protect\citeauthoryear{Levit and Tsoy}{Levit and Tsoy}{2022}]%
        {levit2022theory}
\bibfield{author}{\bibinfo{person}{Doron Levit} {and} \bibinfo{person}{Anton
  Tsoy}.} \bibinfo{year}{2022}\natexlab{}.
\newblock \showarticletitle{A theory of one-size-fits-all recommendations}.
\newblock \bibinfo{journal}{\emph{American Economic Journal: Microeconomics}}
  \bibinfo{volume}{14}, \bibinfo{number}{4} (\bibinfo{year}{2022}),
  \bibinfo{pages}{318--347}.
\newblock


\bibitem[\protect\citeauthoryear{Li, Hu, Guo, Liu, Qi, and Jia}{Li
  et~al\mbox{.}}{2024}]%
        {electronics13040677}
\bibfield{author}{\bibinfo{person}{Shanshan Li}, \bibinfo{person}{Xinzhuan Hu},
  \bibinfo{person}{Jingfeng Guo}, \bibinfo{person}{Bin Liu},
  \bibinfo{person}{Mingyue Qi}, {and} \bibinfo{person}{Yutong Jia}.}
  \bibinfo{year}{2024}\natexlab{}.
\newblock \showarticletitle{Popularity-Debiased Graph Self-Supervised for
  Recommendation}.
\newblock \bibinfo{journal}{\emph{Electronics}} \bibinfo{volume}{13},
  \bibinfo{number}{4} (\bibinfo{year}{2024}).
\newblock
\showISSN{2079-9292}
\urldef\tempurl%
\url{https://doi.org/10.3390/electronics13040677}
\showDOI{\tempurl}


\bibitem[\protect\citeauthoryear{Li, Sun, Ling, and Peng}{Li
  et~al\mbox{.}}{2023}]%
        {li2023survey}
\bibfield{author}{\bibinfo{person}{Xiao Li}, \bibinfo{person}{Li Sun},
  \bibinfo{person}{Mengjie Ling}, {and} \bibinfo{person}{Yan Peng}.}
  \bibinfo{year}{2023}\natexlab{}.
\newblock \showarticletitle{A survey of graph neural network based
  recommendation in social networks}.
\newblock \bibinfo{journal}{\emph{Neurocomputing}}  \bibinfo{volume}{549}
  (\bibinfo{year}{2023}), \bibinfo{pages}{126441}.
\newblock


\bibitem[\protect\citeauthoryear{Lim, Ward, and Benbasat}{Lim
  et~al\mbox{.}}{1997}]%
        {lim1997empirical}
\bibfield{author}{\bibinfo{person}{Kai~H Lim}, \bibinfo{person}{Lawrence~M
  Ward}, {and} \bibinfo{person}{Izak Benbasat}.}
  \bibinfo{year}{1997}\natexlab{}.
\newblock \showarticletitle{An empirical study of computer system learning:
  Comparison of co-discovery and self-discovery methods}.
\newblock \bibinfo{journal}{\emph{Information Systems Research}}
  \bibinfo{volume}{8}, \bibinfo{number}{3} (\bibinfo{year}{1997}),
  \bibinfo{pages}{254--272}.
\newblock


\bibitem[\protect\citeauthoryear{Messinger, Ge, Stroulia, Lyons, Smirnov, and
  Bone}{Messinger et~al\mbox{.}}{2008}]%
        {messinger2008relationship}
\bibfield{author}{\bibinfo{person}{Paul~R Messinger}, \bibinfo{person}{Xin Ge},
  \bibinfo{person}{Eleni Stroulia}, \bibinfo{person}{Kelly Lyons},
  \bibinfo{person}{Kristen Smirnov}, {and} \bibinfo{person}{Michael Bone}.}
  \bibinfo{year}{2008}\natexlab{}.
\newblock \showarticletitle{On the relationship between my avatar and myself}.
\newblock \bibinfo{journal}{\emph{Journal For Virtual Worlds Research}}
  \bibinfo{volume}{1}, \bibinfo{number}{2} (\bibinfo{year}{2008}).
\newblock


\bibitem[\protect\citeauthoryear{Moin}{Moin}{2014}]%
        {moin2014unified}
\bibfield{author}{\bibinfo{person}{Afshin Moin}.}
  \bibinfo{year}{2014}\natexlab{}.
\newblock \showarticletitle{A unified approach to collaborative data
  visualization}. In \bibinfo{booktitle}{\emph{Proceedings of the 29th Annual
  ACM Symposium on Applied Computing}}. \bibinfo{pages}{280--286}.
\newblock


\bibitem[\protect\citeauthoryear{Mu}{Mu}{2018}]%
        {mu2018survey}
\bibfield{author}{\bibinfo{person}{Ruihui Mu}.}
  \bibinfo{year}{2018}\natexlab{}.
\newblock \showarticletitle{A survey of recommender systems based on deep
  learning}.
\newblock \bibinfo{journal}{\emph{Ieee Access}}  \bibinfo{volume}{6}
  (\bibinfo{year}{2018}), \bibinfo{pages}{69009--69022}.
\newblock


\bibitem[\protect\citeauthoryear{Nardi and Harris}{Nardi and Harris}{2006}]%
        {nardi2006strangers}
\bibfield{author}{\bibinfo{person}{Bonnie Nardi} {and} \bibinfo{person}{Justin
  Harris}.} \bibinfo{year}{2006}\natexlab{}.
\newblock \showarticletitle{Strangers and friends: Collaborative play in World
  of Warcraft}. In \bibinfo{booktitle}{\emph{Proceedings of the 2006 20th
  anniversary conference on Computer supported cooperative work}}.
  \bibinfo{pages}{149--158}.
\newblock


\bibitem[\protect\citeauthoryear{Naveed and Ziegler}{Naveed and
  Ziegler}{2020}]%
        {naveed2020featuristic}
\bibfield{author}{\bibinfo{person}{Sidra Naveed} {and}
  \bibinfo{person}{J{\"u}rgen Ziegler}.} \bibinfo{year}{2020}\natexlab{}.
\newblock \showarticletitle{Featuristic: An interactive hybrid system for
  generating explainable recommendations-beyond system accuracy.}. In
  \bibinfo{booktitle}{\emph{IntRS@ RecSys}}. \bibinfo{pages}{14--25}.
\newblock


\bibitem[\protect\citeauthoryear{Noel, Monterola, and Tan}{Noel
  et~al\mbox{.}}{2024}]%
        {noel2024improving}
\bibfield{author}{\bibinfo{person}{Joseph Noel}, \bibinfo{person}{Christopher
  Monterola}, {and} \bibinfo{person}{Daniel~Stanley Tan}.}
  \bibinfo{year}{2024}\natexlab{}.
\newblock \showarticletitle{Improving recommendation diversity without
  retraining from scratch}.
\newblock \bibinfo{journal}{\emph{International Journal of Data Science and
  Analytics}} (\bibinfo{year}{2024}), \bibinfo{pages}{1--10}.
\newblock


\bibitem[\protect\citeauthoryear{O'Donovan, Smyth, Gretarsson, Bostandjiev, and
  H{\"o}llerer}{O'Donovan et~al\mbox{.}}{2008}]%
        {o2008peerchooser}
\bibfield{author}{\bibinfo{person}{John O'Donovan}, \bibinfo{person}{Barry
  Smyth}, \bibinfo{person}{Brynjar Gretarsson}, \bibinfo{person}{Svetlin
  Bostandjiev}, {and} \bibinfo{person}{Tobias H{\"o}llerer}.}
  \bibinfo{year}{2008}\natexlab{}.
\newblock \showarticletitle{PeerChooser: visual interactive recommendation}. In
  \bibinfo{booktitle}{\emph{Proceedings of the SIGCHI Conference on Human
  Factors in Computing Systems}}. \bibinfo{pages}{1085--1088}.
\newblock


\bibitem[\protect\citeauthoryear{Okamoto, Chen, and Li}{Okamoto
  et~al\mbox{.}}{2008}]%
        {okamoto2008ranking}
\bibfield{author}{\bibinfo{person}{Kazuya Okamoto}, \bibinfo{person}{Wei Chen},
  {and} \bibinfo{person}{Xiang-Yang Li}.} \bibinfo{year}{2008}\natexlab{}.
\newblock \showarticletitle{Ranking of closeness centrality for large-scale
  social networks}.
\newblock \bibinfo{journal}{\emph{Lecture Notes in Computer Science}}
  \bibinfo{volume}{5059} (\bibinfo{year}{2008}), \bibinfo{pages}{186--195}.
\newblock


\bibitem[\protect\citeauthoryear{Oyeka, Ebuh, et~al\mbox{.}}{Oyeka
  et~al\mbox{.}}{2012}]%
        {oyeka2012modified}
\bibfield{author}{\bibinfo{person}{Ikewelugo Cyprian~Anaene Oyeka},
  \bibinfo{person}{Godday~Uwawunkonye Ebuh}, {et~al\mbox{.}}}
  \bibinfo{year}{2012}\natexlab{}.
\newblock \showarticletitle{Modified Wilcoxon signed-rank test}.
\newblock \bibinfo{journal}{\emph{Open Journal of Statistics}}
  \bibinfo{volume}{2}, \bibinfo{number}{2} (\bibinfo{year}{2012}),
  \bibinfo{pages}{172--176}.
\newblock


\bibitem[\protect\citeauthoryear{Pariser}{Pariser}{2011}]%
        {pariser2011filter}
\bibfield{author}{\bibinfo{person}{Eli Pariser}.}
  \bibinfo{year}{2011}\natexlab{}.
\newblock \bibinfo{booktitle}{\emph{The filter bubble: What the Internet is
  hiding from you}}.
\newblock \bibinfo{publisher}{penguin UK}.
\newblock


\bibitem[\protect\citeauthoryear{Park and Tuzhilin}{Park and Tuzhilin}{2008}]%
        {park2008long}
\bibfield{author}{\bibinfo{person}{Yoon-Joo Park} {and}
  \bibinfo{person}{Alexander Tuzhilin}.} \bibinfo{year}{2008}\natexlab{}.
\newblock \showarticletitle{The long tail of recommender systems and how to
  leverage it}. In \bibinfo{booktitle}{\emph{Proceedings of the 2008 ACM
  conference on Recommender systems}}. \bibinfo{pages}{11--18}.
\newblock


\bibitem[\protect\citeauthoryear{Parra, Brusilovsky, and Trattner}{Parra
  et~al\mbox{.}}{2014}]%
        {parra2014see}
\bibfield{author}{\bibinfo{person}{Denis Parra}, \bibinfo{person}{Peter
  Brusilovsky}, {and} \bibinfo{person}{Christoph Trattner}.}
  \bibinfo{year}{2014}\natexlab{}.
\newblock \showarticletitle{See what you want to see: visual user-driven
  approach for hybrid recommendation}. In \bibinfo{booktitle}{\emph{Proceedings
  of the 19th international conference on Intelligent User Interfaces}}.
  \bibinfo{pages}{235--240}.
\newblock
\urldef\tempurl%
\url{https://doi.org/10.1145/2557500.2557542}
\showDOI{\tempurl}


\bibitem[\protect\citeauthoryear{Peng}{Peng}{2024}]%
        {peng2024application}
\bibfield{author}{\bibinfo{person}{Yulin Peng}.}
  \bibinfo{year}{2024}\natexlab{}.
\newblock \showarticletitle{The Application of Machine Learning in Predicting
  the Results of Popular eSports Games: Win Rate Prediction in MOBA and FPS
  Games}.
\newblock \bibinfo{journal}{\emph{Highlights in Science, Engineering and
  Technology}}  \bibinfo{volume}{85} (\bibinfo{year}{2024}),
  \bibinfo{pages}{1150--1156}.
\newblock


\bibitem[\protect\citeauthoryear{Pramono, Renalda, and Warnars}{Pramono
  et~al\mbox{.}}{2018}]%
        {pramono2018matchmaking}
\bibfield{author}{\bibinfo{person}{Muhammad~Farrel Pramono},
  \bibinfo{person}{Kevin Renalda}, {and} \bibinfo{person}{Harco Leslie
  Hendric~Spits Warnars}.} \bibinfo{year}{2018}\natexlab{}.
\newblock \showarticletitle{Matchmaking problems in MOBA Games}.
\newblock \bibinfo{journal}{\emph{Indonesian Journal of Electrical Engineering
  and Computer Science}} \bibinfo{volume}{11}, \bibinfo{number}{3}
  (\bibinfo{year}{2018}), \bibinfo{pages}{908--917}.
\newblock


\bibitem[\protect\citeauthoryear{Radford, Kim, Hallacy, Ramesh, Goh, Agarwal,
  Sastry, Askell, Mishkin, Clark, et~al\mbox{.}}{Radford et~al\mbox{.}}{2021}]%
        {radford2021learning}
\bibfield{author}{\bibinfo{person}{Alec Radford}, \bibinfo{person}{Jong~Wook
  Kim}, \bibinfo{person}{Chris Hallacy}, \bibinfo{person}{Aditya Ramesh},
  \bibinfo{person}{Gabriel Goh}, \bibinfo{person}{Sandhini Agarwal},
  \bibinfo{person}{Girish Sastry}, \bibinfo{person}{Amanda Askell},
  \bibinfo{person}{Pamela Mishkin}, \bibinfo{person}{Jack Clark},
  {et~al\mbox{.}}} \bibinfo{year}{2021}\natexlab{}.
\newblock \showarticletitle{Learning transferable visual models from natural
  language supervision}. In \bibinfo{booktitle}{\emph{International conference
  on machine learning}}. PMLR, \bibinfo{pages}{8748--8763}.
\newblock


\bibitem[\protect\citeauthoryear{Ramirez}{Ramirez}{2018}]%
        {ramirez2018good}
\bibfield{author}{\bibinfo{person}{Fanny~Anne Ramirez}.}
  \bibinfo{year}{2018}\natexlab{}.
\newblock \showarticletitle{From good associates to true friends: An
  exploration of friendship practices in massively multiplayer online games}.
\newblock \bibinfo{journal}{\emph{Social interactions in virtual worlds: An
  interdisciplinary perspective}} (\bibinfo{year}{2018}),
  \bibinfo{pages}{62--79}.
\newblock


\bibitem[\protect\citeauthoryear{Raza, Rahman, Kamawal, Toroghi, Raval, Navah,
  and Kazemeini}{Raza et~al\mbox{.}}{2024}]%
        {raza2024comprehensive}
\bibfield{author}{\bibinfo{person}{Shaina Raza}, \bibinfo{person}{Mizanur
  Rahman}, \bibinfo{person}{Safiullah Kamawal}, \bibinfo{person}{Armin
  Toroghi}, \bibinfo{person}{Ananya Raval}, \bibinfo{person}{Farshad Navah},
  {and} \bibinfo{person}{Amirmohammad Kazemeini}.}
  \bibinfo{year}{2024}\natexlab{}.
\newblock \showarticletitle{A Comprehensive Review of Recommender Systems:
  Transitioning from Theory to Practice}.
\newblock \bibinfo{journal}{\emph{arXiv preprint arXiv:2407.13699}}
  (\bibinfo{year}{2024}).
\newblock


\bibitem[\protect\citeauthoryear{Riyadh, Arya, Chan, and Imran}{Riyadh
  et~al\mbox{.}}{2020}]%
        {riyadh2020enhancing}
\bibfield{author}{\bibinfo{person}{Md Riyadh}, \bibinfo{person}{Ali Arya},
  \bibinfo{person}{Gerry Chan}, {and} \bibinfo{person}{Masud Imran}.}
  \bibinfo{year}{2020}\natexlab{}.
\newblock \showarticletitle{Enhancing Social Ties Through Manual Player
  Matchmaking in Online Multiplayer Games}. In \bibinfo{booktitle}{\emph{HCI
  International 2020--Late Breaking Papers: Cognition, Learning and Games: 22nd
  HCI International Conference, HCII 2020, Copenhagen, Denmark, July 19--24,
  2020, Proceedings 22}}. Springer, \bibinfo{pages}{708--729}.
\newblock
\urldef\tempurl%
\url{https://doi.org/10.1007/978-3-030-60128-7_52}
\showDOI{\tempurl}


\bibitem[\protect\citeauthoryear{Schaekermann}{Schaekermann}{2020}]%
        {schaekermann2020human}
\bibfield{author}{\bibinfo{person}{Mike Schaekermann}.}
  \bibinfo{year}{2020}\natexlab{}.
\newblock \showarticletitle{Human-ai interaction in the presence of ambiguity:
  From deliberation-based labeling to ambiguity-aware ai}.
\newblock  (\bibinfo{year}{2020}).
\newblock


\bibitem[\protect\citeauthoryear{Seo, Kim, Lee, and Baik}{Seo
  et~al\mbox{.}}{2017}]%
        {seo2017personalized}
\bibfield{author}{\bibinfo{person}{Young-Duk Seo}, \bibinfo{person}{Young-Gab
  Kim}, \bibinfo{person}{Euijong Lee}, {and} \bibinfo{person}{Doo-Kwon Baik}.}
  \bibinfo{year}{2017}\natexlab{}.
\newblock \showarticletitle{Personalized recommender system based on friendship
  strength in social network services}.
\newblock \bibinfo{journal}{\emph{Expert Systems with Applications}}
  \bibinfo{volume}{69} (\bibinfo{year}{2017}), \bibinfo{pages}{135--148}.
\newblock


\bibitem[\protect\citeauthoryear{Shi}{Shi}{2013}]%
        {shi2013trading}
\bibfield{author}{\bibinfo{person}{Lei Shi}.} \bibinfo{year}{2013}\natexlab{}.
\newblock \showarticletitle{Trading-off among accuracy, similarity, diversity,
  and long-tail: a graph-based recommendation approach}. In
  \bibinfo{booktitle}{\emph{Proceedings of the 7th ACM Conference on
  Recommender Systems}}. \bibinfo{pages}{57--64}.
\newblock


\bibitem[\protect\citeauthoryear{Shneiderman}{Shneiderman}{2003}]%
        {shneiderman2003eyes}
\bibfield{author}{\bibinfo{person}{Ben Shneiderman}.}
  \bibinfo{year}{2003}\natexlab{}.
\newblock \showarticletitle{The eyes have it: A task by data type taxonomy for
  information visualizations}.
\newblock In \bibinfo{booktitle}{\emph{The craft of information
  visualization}}. \bibinfo{publisher}{Elsevier}, \bibinfo{pages}{364--371}.
\newblock


\bibitem[\protect\citeauthoryear{Sun, Lank, and Terry}{Sun
  et~al\mbox{.}}{2017}]%
        {sun2017label}
\bibfield{author}{\bibinfo{person}{Yunjia Sun}, \bibinfo{person}{Edward Lank},
  {and} \bibinfo{person}{Michael Terry}.} \bibinfo{year}{2017}\natexlab{}.
\newblock \showarticletitle{Label-and-learn: Visualizing the likelihood of
  machine learning classifier's success during data labeling}. In
  \bibinfo{booktitle}{\emph{Proceedings of the 22nd International Conference on
  Intelligent User Interfaces}}. \bibinfo{pages}{523--534}.
\newblock


\bibitem[\protect\citeauthoryear{Tian, Wang, Xie, Ma, and Li}{Tian
  et~al\mbox{.}}{2024}]%
        {10.1145/3565698.3565765}
\bibfield{author}{\bibinfo{person}{Yun Tian}, \bibinfo{person}{He Wang},
  \bibinfo{person}{Laixin Xie}, \bibinfo{person}{Xiaojuan Ma}, {and}
  \bibinfo{person}{Quan Li}.} \bibinfo{year}{2024}\natexlab{}.
\newblock \showarticletitle{VFLens: Co-design the Modeling Process for
  Efficient Vertical Federated Learning via Visualization}. In
  \bibinfo{booktitle}{\emph{Proceedings of the Tenth International Symposium of
  Chinese CHI}} (<conf-loc>, <city>Guangzhou, China and Online</city>,
  <country>China</country>, </conf-loc>) \emph{(\bibinfo{series}{Chinese CHI
  '22})}. \bibinfo{publisher}{Association for Computing Machinery},
  \bibinfo{address}{New York, NY, USA}, \bibinfo{pages}{1–14}.
\newblock
\showISBNx{9781450398695}
\urldef\tempurl%
\url{https://doi.org/10.1145/3565698.3565765}
\showDOI{\tempurl}


\bibitem[\protect\citeauthoryear{Tsai and Brusilovsky}{Tsai and
  Brusilovsky}{2017}]%
        {tsai2017leveraging}
\bibfield{author}{\bibinfo{person}{Chun-Hua Tsai} {and} \bibinfo{person}{Peter
  Brusilovsky}.} \bibinfo{year}{2017}\natexlab{}.
\newblock \showarticletitle{Leveraging interfaces to improve recommendation
  diversity}. In \bibinfo{booktitle}{\emph{Adjunct Publication of the 25th
  Conference on User Modeling, Adaptation and Personalization}}.
  \bibinfo{pages}{65--70}.
\newblock
\urldef\tempurl%
\url{https://doi.org/10.1145/3099023.3099073}
\showDOI{\tempurl}


\bibitem[\protect\citeauthoryear{Vahlo, Kaakinen, Holm, and Koponen}{Vahlo
  et~al\mbox{.}}{2017}]%
        {vahlo2017digital}
\bibfield{author}{\bibinfo{person}{Jukka Vahlo}, \bibinfo{person}{Johanna~K
  Kaakinen}, \bibinfo{person}{Suvi~K Holm}, {and} \bibinfo{person}{Aki
  Koponen}.} \bibinfo{year}{2017}\natexlab{}.
\newblock \showarticletitle{Digital game dynamics preferences and player
  types}.
\newblock \bibinfo{journal}{\emph{Journal of Computer-Mediated Communication}}
  \bibinfo{volume}{22}, \bibinfo{number}{2} (\bibinfo{year}{2017}),
  \bibinfo{pages}{88--103}.
\newblock


\bibitem[\protect\citeauthoryear{Van~der Maaten and Hinton}{Van~der Maaten and
  Hinton}{2008}]%
        {van2008visualizing}
\bibfield{author}{\bibinfo{person}{Laurens Van~der Maaten} {and}
  \bibinfo{person}{Geoffrey Hinton}.} \bibinfo{year}{2008}\natexlab{}.
\newblock \showarticletitle{Visualizing data using t-SNE.}
\newblock \bibinfo{journal}{\emph{Journal of machine learning research}}
  \bibinfo{volume}{9}, \bibinfo{number}{11} (\bibinfo{year}{2008}).
\newblock


\bibitem[\protect\citeauthoryear{Vella, Klarkowski, Turkay, and Johnson}{Vella
  et~al\mbox{.}}{2020}]%
        {vella2020making}
\bibfield{author}{\bibinfo{person}{Kellie Vella}, \bibinfo{person}{Madison
  Klarkowski}, \bibinfo{person}{Selen Turkay}, {and} \bibinfo{person}{Daniel
  Johnson}.} \bibinfo{year}{2020}\natexlab{}.
\newblock \showarticletitle{Making friends in online games: gender differences
  and designing for greater social connectedness}.
\newblock \bibinfo{journal}{\emph{Behaviour \& Information Technology}}
  \bibinfo{volume}{39}, \bibinfo{number}{8} (\bibinfo{year}{2020}),
  \bibinfo{pages}{917--934}.
\newblock


\bibitem[\protect\citeauthoryear{Verbert, Parra, and Brusilovsky}{Verbert
  et~al\mbox{.}}{2014}]%
        {verbert2014effect}
\bibfield{author}{\bibinfo{person}{Katrien Verbert}, \bibinfo{person}{Denis
  Parra}, {and} \bibinfo{person}{Peter Brusilovsky}.}
  \bibinfo{year}{2014}\natexlab{}.
\newblock \showarticletitle{The effect of different set-based visualizations on
  user exploration of recommendations}. In \bibinfo{booktitle}{\emph{CEUR
  Workshop Proceedings}}, Vol.~\bibinfo{volume}{1253}. University of
  Pittsburgh, \bibinfo{pages}{37--44}.
\newblock


\bibitem[\protect\citeauthoryear{Verbert, Parra, Brusilovsky, and
  Duval}{Verbert et~al\mbox{.}}{2013}]%
        {verbert2013visualizing}
\bibfield{author}{\bibinfo{person}{Katrien Verbert}, \bibinfo{person}{Denis
  Parra}, \bibinfo{person}{Peter Brusilovsky}, {and} \bibinfo{person}{Erik
  Duval}.} \bibinfo{year}{2013}\natexlab{}.
\newblock \showarticletitle{Visualizing recommendations to support exploration,
  transparency and controllability}. In \bibinfo{booktitle}{\emph{Proceedings
  of the 2013 international conference on Intelligent user interfaces}}.
  \bibinfo{pages}{351--362}.
\newblock


\bibitem[\protect\citeauthoryear{Wang, Ouyang, Wu, Jiang, Jin, Cao, and
  Li}{Wang et~al\mbox{.}}{2024}]%
        {wang2024kmtlabeler}
\bibfield{author}{\bibinfo{person}{He Wang}, \bibinfo{person}{Yang Ouyang},
  \bibinfo{person}{Yuchen Wu}, \bibinfo{person}{Chang Jiang},
  \bibinfo{person}{Lixia Jin}, \bibinfo{person}{Yuanwu Cao}, {and}
  \bibinfo{person}{Quan Li}.} \bibinfo{year}{2024}\natexlab{}.
\newblock \showarticletitle{KMTLabeler: An Interactive Knowledge-Assisted
  Labeling Tool for Medical Text Classification}.
\newblock \bibinfo{journal}{\emph{IEEE Transactions on Visualization and
  Computer Graphics}} (\bibinfo{year}{2024}).
\newblock


\bibitem[\protect\citeauthoryear{Wang, Liao, Cao, Qi, and Wang}{Wang
  et~al\mbox{.}}{2014}]%
        {wang2014friendbook}
\bibfield{author}{\bibinfo{person}{Zhibo Wang}, \bibinfo{person}{Jilong Liao},
  \bibinfo{person}{Qing Cao}, \bibinfo{person}{Hairong Qi}, {and}
  \bibinfo{person}{Zhi Wang}.} \bibinfo{year}{2014}\natexlab{}.
\newblock \showarticletitle{Friendbook: a semantic-based friend recommendation
  system for social networks}.
\newblock \bibinfo{journal}{\emph{IEEE transactions on mobile computing}}
  \bibinfo{volume}{14}, \bibinfo{number}{3} (\bibinfo{year}{2014}),
  \bibinfo{pages}{538--551}.
\newblock
\urldef\tempurl%
\url{https://doi.org/10.1109/tmc.2014.2322373}
\showDOI{\tempurl}


\bibitem[\protect\citeauthoryear{Ward}{Ward}{2022}]%
        {ward2022network}
\bibfield{author}{\bibinfo{person}{Michael~R Ward}.}
  \bibinfo{year}{2022}\natexlab{}.
\newblock \showarticletitle{Network engagement from learning friends’
  preferences: evidence from a video gaming social network}.
\newblock \bibinfo{journal}{\emph{Electronic Markets}} \bibinfo{volume}{32},
  \bibinfo{number}{3} (\bibinfo{year}{2022}), \bibinfo{pages}{1239--1255}.
\newblock


\bibitem[\protect\citeauthoryear{Watson, Watson, and Zheng}{Watson
  et~al\mbox{.}}{2019}]%
        {watson2019study}
\bibfield{author}{\bibinfo{person}{Bryan Watson}, \bibinfo{person}{Thomas
  Watson}, {and} \bibinfo{person}{Jun Zheng}.} \bibinfo{year}{2019}\natexlab{}.
\newblock \showarticletitle{A study of friend recommendations for gaming
  communities}.
\newblock \bibinfo{journal}{\emph{International Journal of Web Based
  Communities}} \bibinfo{volume}{15}, \bibinfo{number}{4}
  (\bibinfo{year}{2019}), \bibinfo{pages}{292--314}.
\newblock
\urldef\tempurl%
\url{https://doi.org/10.1504/ijwbc.2019.10023720}
\showDOI{\tempurl}


\bibitem[\protect\citeauthoryear{Wu, Sun, Zhang, Xie, and Cui}{Wu
  et~al\mbox{.}}{2022}]%
        {wu2022graph}
\bibfield{author}{\bibinfo{person}{Shiwen Wu}, \bibinfo{person}{Fei Sun},
  \bibinfo{person}{Wentao Zhang}, \bibinfo{person}{Xu Xie}, {and}
  \bibinfo{person}{Bin Cui}.} \bibinfo{year}{2022}\natexlab{}.
\newblock \showarticletitle{Graph neural networks in recommender systems: a
  survey}.
\newblock \bibinfo{journal}{\emph{Comput. Surveys}} \bibinfo{volume}{55},
  \bibinfo{number}{5} (\bibinfo{year}{2022}), \bibinfo{pages}{1--37}.
\newblock


\bibitem[\protect\citeauthoryear{Xie, Liu, Liu, Zhang, Cui, Zhang, and Lin}{Xie
  et~al\mbox{.}}{2021}]%
        {xie2021improving}
\bibfield{author}{\bibinfo{person}{Ruobing Xie}, \bibinfo{person}{Qi Liu},
  \bibinfo{person}{Shukai Liu}, \bibinfo{person}{Ziwei Zhang},
  \bibinfo{person}{Peng Cui}, \bibinfo{person}{Bo Zhang}, {and}
  \bibinfo{person}{Leyu Lin}.} \bibinfo{year}{2021}\natexlab{}.
\newblock \showarticletitle{Improving accuracy and diversity in matching of
  recommendation with diversified preference network}.
\newblock \bibinfo{journal}{\emph{IEEE Transactions on Big Data}}
  \bibinfo{volume}{8}, \bibinfo{number}{4} (\bibinfo{year}{2021}),
  \bibinfo{pages}{955--967}.
\newblock
\urldef\tempurl%
\url{https://doi.org/10.1109/tbdata.2021.3103263}
\showDOI{\tempurl}


\bibitem[\protect\citeauthoryear{Xie, Qiu, Rao, Liu, Zhang, and Lin}{Xie
  et~al\mbox{.}}{2020}]%
        {xie2020internal}
\bibfield{author}{\bibinfo{person}{Ruobing Xie}, \bibinfo{person}{Zhijie Qiu},
  \bibinfo{person}{Jun Rao}, \bibinfo{person}{Yi Liu}, \bibinfo{person}{Bo
  Zhang}, {and} \bibinfo{person}{Leyu Lin}.} \bibinfo{year}{2020}\natexlab{}.
\newblock \showarticletitle{Internal and Contextual Attention Network for
  Cold-start Multi-channel Matching in Recommendation.}. In
  \bibinfo{booktitle}{\emph{IJCAI}}. \bibinfo{pages}{2732--2738}.
\newblock
\urldef\tempurl%
\url{https://doi.org/10.24963/ijcai.2020/379}
\showDOI{\tempurl}


\bibitem[\protect\citeauthoryear{Xu, Zhou, and Ma}{Xu et~al\mbox{.}}{2019}]%
        {xu2019scholar}
\bibfield{author}{\bibinfo{person}{Yunhong Xu}, \bibinfo{person}{Duanning
  Zhou}, {and} \bibinfo{person}{Jian Ma}.} \bibinfo{year}{2019}\natexlab{}.
\newblock \showarticletitle{Scholar-friend recommendation in online academic
  communities: an approach based on heterogeneous network}.
\newblock \bibinfo{journal}{\emph{Decision Support Systems}}
  \bibinfo{volume}{119} (\bibinfo{year}{2019}), \bibinfo{pages}{1--13}.
\newblock
\urldef\tempurl%
\url{https://doi.org/10.1016/j.dss.2019.01.004}
\showDOI{\tempurl}


\bibitem[\protect\citeauthoryear{Yang, Liu, Wang, Wang, Fan, and Yu}{Yang
  et~al\mbox{.}}{2022}]%
        {yang2022large}
\bibfield{author}{\bibinfo{person}{Liangwei Yang}, \bibinfo{person}{Zhiwei
  Liu}, \bibinfo{person}{Yu Wang}, \bibinfo{person}{Chen Wang},
  \bibinfo{person}{Ziwei Fan}, {and} \bibinfo{person}{Philip~S Yu}.}
  \bibinfo{year}{2022}\natexlab{}.
\newblock \showarticletitle{Large-scale personalized video game recommendation
  via social-aware contextualized graph neural network}. In
  \bibinfo{booktitle}{\emph{Proceedings of the ACM Web Conference 2022}}.
  \bibinfo{pages}{3376--3386}.
\newblock


\bibitem[\protect\citeauthoryear{Ye and Ji}{Ye and Ji}{2021}]%
        {ye2021sparse}
\bibfield{author}{\bibinfo{person}{Yang Ye} {and} \bibinfo{person}{Shihao Ji}.}
  \bibinfo{year}{2021}\natexlab{}.
\newblock \showarticletitle{Sparse graph attention networks}.
\newblock \bibinfo{journal}{\emph{IEEE Transactions on Knowledge and Data
  Engineering}} \bibinfo{volume}{35}, \bibinfo{number}{1}
  (\bibinfo{year}{2021}), \bibinfo{pages}{905--916}.
\newblock


\bibitem[\protect\citeauthoryear{Yin, Fang, Chen, and Sheng}{Yin
  et~al\mbox{.}}{2022}]%
        {Yin2022DiversityPL}
\bibfield{author}{\bibinfo{person}{Kexin Yin}, \bibinfo{person}{Xiao Fang},
  \bibinfo{person}{Bi-Yi Chen}, {and} \bibinfo{person}{Olivia R.~Liu Sheng}.}
  \bibinfo{year}{2022}\natexlab{}.
\newblock \showarticletitle{Diversity Preference-Aware Link Recommendation for
  Online Social Networks}.
\newblock \bibinfo{journal}{\emph{ArXiv}}  \bibinfo{volume}{abs/2205.10689}
  (\bibinfo{year}{2022}).
\newblock
\urldef\tempurl%
\url{https://doi.org/10.2139/ssrn.4158074}
\showDOI{\tempurl}


\bibitem[\protect\citeauthoryear{Zhang, He, Wu, Huang, Qin, Wang, Ye, Huang,
  Liao, Chen, et~al\mbox{.}}{Zhang et~al\mbox{.}}{2023}]%
        {zhang2023pathnarratives}
\bibfield{author}{\bibinfo{person}{Heyu Zhang}, \bibinfo{person}{Yan He},
  \bibinfo{person}{Xiaomin Wu}, \bibinfo{person}{Peixiang Huang},
  \bibinfo{person}{Wenkang Qin}, \bibinfo{person}{Fan Wang},
  \bibinfo{person}{Juxiang Ye}, \bibinfo{person}{Xirui Huang},
  \bibinfo{person}{Yanfang Liao}, \bibinfo{person}{Hang Chen}, {et~al\mbox{.}}}
  \bibinfo{year}{2023}\natexlab{}.
\newblock \showarticletitle{PathNarratives: Data annotation for pathological
  human-AI collaborative diagnosis}.
\newblock \bibinfo{journal}{\emph{Frontiers in Medicine}}  \bibinfo{volume}{9}
  (\bibinfo{year}{2023}), \bibinfo{pages}{1070072}.
\newblock


\bibitem[\protect\citeauthoryear{Zhang and Guo}{Zhang and Guo}{2006}]%
        {zhang2006method}
\bibfield{author}{\bibinfo{person}{Li Zhang} {and} \bibinfo{person}{Jun Guo}.}
  \bibinfo{year}{2006}\natexlab{}.
\newblock \showarticletitle{A method for the selection of training samples
  based on boundary samples}.
\newblock \bibinfo{journal}{\emph{Journal of Beijing University of Posts and
  Telecommunications}} \bibinfo{volume}{29}, \bibinfo{number}{4}
  (\bibinfo{year}{2006}), \bibinfo{pages}{77}.
\newblock


\bibitem[\protect\citeauthoryear{Zhang and Li}{Zhang and Li}{2022a}]%
        {Zhang2022ALF}
\bibfield{author}{\bibinfo{person}{Lin Zhang} {and} \bibinfo{person}{Rui Li}.}
  \bibinfo{year}{2022}\natexlab{a}.
\newblock \showarticletitle{A Large-scale Friend Suggestion Architecture}.
\newblock \bibinfo{journal}{\emph{ArXiv}}  \bibinfo{volume}{abs/2212.12773}
  (\bibinfo{year}{2022}).
\newblock
\urldef\tempurl%
\url{https://doi.org/10.1109/dsaa54385.2022.10032355}
\showDOI{\tempurl}


\bibitem[\protect\citeauthoryear{Zhang and Li}{Zhang and Li}{2022b}]%
        {zhang2022large}
\bibfield{author}{\bibinfo{person}{Lin Zhang} {and} \bibinfo{person}{Rui Li}.}
  \bibinfo{year}{2022}\natexlab{b}.
\newblock \showarticletitle{A Large-scale Friend Suggestion Architecture}.
\newblock \bibinfo{journal}{\emph{arXiv preprint arXiv:2212.12773}}
  (\bibinfo{year}{2022}).
\newblock
\urldef\tempurl%
\url{https://doi.org/10.1109/dsaa54385.2022.10032355}
\showDOI{\tempurl}


\bibitem[\protect\citeauthoryear{Zhang, Wei, Zhang, Wang, and Ho}{Zhang
  et~al\mbox{.}}{2020}]%
        {zhang2020diversity}
\bibfield{author}{\bibinfo{person}{Liang Zhang}, \bibinfo{person}{Quanshen
  Wei}, \bibinfo{person}{Lei Zhang}, \bibinfo{person}{Baojiao Wang}, {and}
  \bibinfo{person}{Wen-Hsien Ho}.} \bibinfo{year}{2020}\natexlab{}.
\newblock \showarticletitle{Diversity balancing for two-stage collaborative
  filtering in recommender systems}.
\newblock \bibinfo{journal}{\emph{Applied Sciences}} \bibinfo{volume}{10},
  \bibinfo{number}{4} (\bibinfo{year}{2020}), \bibinfo{pages}{1257}.
\newblock


\bibitem[\protect\citeauthoryear{Zhang, Sun, Lin, Xiao, and Tang}{Zhang
  et~al\mbox{.}}{2022a}]%
        {zhang2022measuring}
\bibfield{author}{\bibinfo{person}{Shiqi Zhang}, \bibinfo{person}{Jiachen Sun},
  \bibinfo{person}{Wenqing Lin}, \bibinfo{person}{Xiaokui Xiao}, {and}
  \bibinfo{person}{Bo Tang}.} \bibinfo{year}{2022}\natexlab{a}.
\newblock \showarticletitle{Measuring Friendship Closeness: A Perspective of
  Social Identity Theory}.
\newblock \bibinfo{journal}{\emph{arXiv preprint arXiv:2208.09176}}
  (\bibinfo{year}{2022}).
\newblock
\urldef\tempurl%
\url{https://doi.org/10.1145/3511808.3557076}
\showDOI{\tempurl}


\bibitem[\protect\citeauthoryear{Zhang, Wang, Zhang, Zhu, Chen, and
  Zhang}{Zhang et~al\mbox{.}}{2022b}]%
        {zhang2022onelabeler}
\bibfield{author}{\bibinfo{person}{Yu Zhang}, \bibinfo{person}{Yun Wang},
  \bibinfo{person}{Haidong Zhang}, \bibinfo{person}{Bin Zhu},
  \bibinfo{person}{Siming Chen}, {and} \bibinfo{person}{Dongmei Zhang}.}
  \bibinfo{year}{2022}\natexlab{b}.
\newblock \showarticletitle{Onelabeler: A flexible system for building data
  labeling tools}. In \bibinfo{booktitle}{\emph{Proceedings of the 2022 CHI
  conference on human factors in computing systems}}. \bibinfo{pages}{1--22}.
\newblock


\bibitem[\protect\citeauthoryear{Zhao, Wang, Liu, Cheng, Aggarwal, and
  Derr}{Zhao et~al\mbox{.}}{2023}]%
        {zhao2023fairness}
\bibfield{author}{\bibinfo{person}{Yuying Zhao}, \bibinfo{person}{Yu Wang},
  \bibinfo{person}{Yunchao Liu}, \bibinfo{person}{Xueqi Cheng},
  \bibinfo{person}{Charu~C Aggarwal}, {and} \bibinfo{person}{Tyler Derr}.}
  \bibinfo{year}{2023}\natexlab{}.
\newblock \showarticletitle{Fairness and diversity in recommender systems: a
  survey}.
\newblock \bibinfo{journal}{\emph{ACM Transactions on Intelligent Systems and
  Technology}} (\bibinfo{year}{2023}).
\newblock


\bibitem[\protect\citeauthoryear{Zhou and Han}{Zhou and Han}{2019}]%
        {zhou2019personalized}
\bibfield{author}{\bibinfo{person}{Wen Zhou} {and} \bibinfo{person}{Wenbo
  Han}.} \bibinfo{year}{2019}\natexlab{}.
\newblock \showarticletitle{Personalized recommendation via user preference
  matching}.
\newblock \bibinfo{journal}{\emph{Information Processing \& Management}}
  \bibinfo{volume}{56}, \bibinfo{number}{3} (\bibinfo{year}{2019}),
  \bibinfo{pages}{955--968}.
\newblock


\end{thebibliography}

\appendix
\section{Design Alternative of Preference Projection View}
\label{hexbin_alter}
\begin{figure}[h]
\centering 
\includegraphics[width=\linewidth]{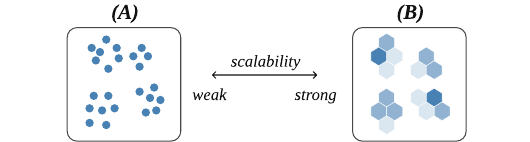}
\caption{An alternative design of hexbin plot.}
\label{fig:hexbin}
\end{figure}

\par An alternative design involves employing a scatter plot (\cref{fig:hexbin}(A)), but it may encounter performance issues when dealing with a large number of players. The rationale behind selecting the hexbin plot (\cref{fig:hexbin}(B)) is its robust scalability, guaranteeing a seamless operation even with a substantial volume of player data.

\section{Design Alternative of LineUp}
\label{lineup_alter}
\begin{figure}[h]
\centering 
\includegraphics[width=\linewidth]{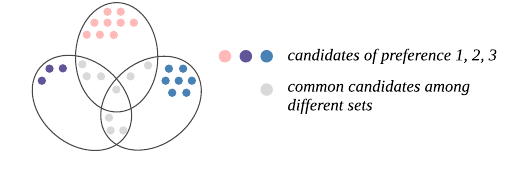}
\caption{An alternative design of Lineup View.}
\label{fig:venn}
\end{figure}

\par We proposed an alternative design for the candidate generation process that combines a Venn diagram and a recommendation list (\cref{fig:venn}). The Venn diagram visualizes each set of generated candidates based on a specific aspect, with overlapping areas representing common candidates among different sets. The recommendation list displays each candidate colored consistently with its corresponding source in the Venn diagram. However, this alternative design can result in a complex color-coding scheme, especially when the number of sets in the Venn diagram is large, making it challenging to efficiently compare and interpret the recommendations. The lineup visualization method allows for more intuitive and efficient inspection and comparison of candidates from various aspects.

\end{document}